# The NonHuman Primate Neuroimaging & Neuroanatomy Project

**Authors**
Takuya Hayashi[1,2]*, Yujie Hou[3†], Matthew F Glasser[4,5†], Joonas A Autio[1†], Kenneth Knoblauch[3], Miho Inoue-Murayama[6], Tim Coalson[4], Essa Yacoub[7], Stephen Smith[8], Henry Kennedy[3,9‡], and David C Van Essen[4‡]

**Affiliations**
[1]RIKEN Center for Biosystems Dynamics Research, Kobe, Japan
[2]Department of Neurobiology, Kyoto University Graduate School of Medicine, Kyoto, Japan
[3]Univ Lyon, Université Claude Bernard Lyon 1, Inserm, Stem Cell and Brain Research Institute U1208, Bron, France
[4]Department of Neuroscience and [5]Radiology, Washington University Medical School, St Louis, MO USA
[6]Wildlife Research Center, Kyoto University, Kyoto, Japan
[7]Center for Magnetic Resonance Research, Department of Radiology, University of Minnesota, Minneapolis, USA
[8]Oxford Centre for Functional Magnetic Resonance Imaging of the Brain (FMRIB), Wellcome Centre for Integrative Neuroimaging (WIN), Nuffield Department of Clinical Neurosciences, Oxford University, Oxford, UK
[9]Institute of Neuroscience, State Key Laboratory of Neuroscience, Chinese Academy of Sciences (CAS) Key Laboratory of Primate Neurobiology, CAS, Shanghai, China

[†‡]**Equal contributions**

*Corresponding author
Takuya Hayashi
Laboratory for Brain Connectomics Imaging,
RIKEN Center for Biosystems Dynamics Research
6-7-3 MI R&D Center 3F, Minatojima-minamimachi,
Chuo-ku, Kobe 650-0047, Japan

**Author contributions**
Takuya Hayashi: Conceptualization, Funding acquisition, Investigation, Formal Analysis, Writing - original draft, review & editing
Yujie Hou: Investigation, Formal Analysis, Writing - review and editing
Matthew F. Glasser: Conceptualization, Investigation, Formal Analysis, Writing - review and editing
Kenneth Knoblauch: Investigation, Formal Analysis
Miho Inoue-Murayama: Investigation, Formal Analysis
Joonas A. Autio: Investigation, Formal Analysis
Tim Coalson: Investigation, Formal Analysis, Writing - review & editing
Essa Yacoub: Investigation, Writing - review & editing
Stephen Smith: Investigation, Writing - review & editing
Henry Kennedy: Conceptualization, Funding acquisition, Investigation, Writing - original draft, review & editing
David C. Van Essen: Conceptualization, Funding acquisition, Investigation, Writing - original draft, review & editing

**Acknowledgements**
This study is supported by a grant Brain/MINDS-beyond from Japan Agency of Medical Research and Development (AMED) (JP20dm0307006h0002) (T.H., M.I.M.), grant KAKENHI from MEXT (19H04904) (M.I.M.), NIH Grant MH060974 (D.C.V.E.), NIH Grant P50NS098573 (E.Y.) and Wellcome Trust (S.M.S.). The grants from SBRI are LABEX CORTEX ANR-11-LABX-0042; Université de Lyon ANR-11-IDEX-0007, A2P2MC ANR-17-NEUC-0004, CORTICITY ANR-17-HBPR-0003, Région Auvergne-Rhône-Alpes SCUSI 1700933701 (H.K.) and DUAL_STREAMS ANR-19-CE37-0025 (K.K.). The authors appreciate technical contributions from Yuki Hori, Atsushi Yoshida, Kantaro Nishigori, Chihiro Yokoyama, Takayuki Ose, Masahiro Ohno, Chihiro Takeda, Akihiro Kawasaki, Kenji Mitsui, Sumika Sagawa, Reiko Kobayashi, Takuro Ikeda, Toshihiko Aso, Yuki Matsumoto, Takashi Azuma, Masahiko Takada, Chad Donahue, John Harwell, Erin Reid.

**Abstract**
Multi-modal neuroimaging projects such as the Human Connectome Project (HCP) and UK Biobank are advancing our understanding of human brain architecture, function, connectivity, and their variability across individuals using high-quality non-invasive data from many subjects. Such efforts depend upon the accuracy of non-invasive brain imaging measures. However, 'ground truth' validation of connectivity using invasive tracers is not feasible in humans. Studies using nonhuman primates (NHPs) enable comparisons between invasive and non-invasive measures, including exploration of how "functional connectivity" from fMRI and "tractographic connectivity" from diffusion MRI compare with long-distance connections measured using tract tracing. Our NonHuman Primate Neuroimaging & Neuroanatomy Project (NHP_NNP) is an international effort (6 laboratories in 5 countries) to: (i) acquire and analyze high-quality multi-modal brain imaging data of macaque and marmoset monkeys using protocols and methods adapted from the HCP; (ii) acquire quantitative invasive tract-tracing data for cortical and subcortical projections to cortical areas; and (iii) map the distributions of different brain cell types with immunocytochemical stains to better define brain areal boundaries. We are acquiring high-resolution structural, functional, and diffusion MRI data together with behavioral measures from over 100 individual macaques and marmosets in order to generate non-invasive measures of brain architecture such as myelin and cortical thickness maps, as well as functional and diffusion tractography-based connectomes. We are using classical and next-generation anatomical tracers to generate quantitative connectivity maps based on brain-wide counting of labeled cortical and subcortical neurons, providing ground truth measures of connectivity. Advanced statistical modeling techniques address the consistency of both kinds of data across individuals, allowing comparison of tracer-based and non-invasive MRI-based connectivity measures. We aim to develop improved cortical and subcortical areal atlases by combining histological and imaging methods. Finally, we are collecting genetic and sociality-associated behavioral data in all animals in an effort to understand how genetic variation shapes the connectome and behavior.

**Abbreviations**
CMRR: Center for Magnetic Resonance Research (at UMinn)
CSF: cerebrospinal fluid
dMRI: diffusion-weighted MRI
DMN: default mode network
DTe: ex vivo diffusion based tractography
DTi: in vivo diffusion based tractography
EDR: exponential distance rule
EPI: echo planar imaging
FLNe: fraction of labeled neurons
fMRI: functional MRI
FSe: Fractional Scaling of streamlines
GCA: Gaussian Classifier Atlas
HCP: human connectome project
IoN: Institute of Neuroscience, Shanghai University, Shanghai, China
KU: Kyoto University
mCT: median cortical thickness
MRI: magnetic resonance imaging
NHP: nonhuman primates
NNP: neuroimaging and neuroanatomy project
OU: Oxford University, Oxford, UK
RIKEN: Institute of Physical and Chemical Research, Japan
SBRI: Stem cell Brain Research Institute, Lyon, France
SNR: signal to noise ratio

TT: tract tracing
UMinn: University of Minnesota, USA
YA-HCP: young adult human connectome project
WUSTL: Washington University in St. Louis, MO USA

# 1. Introduction

For over a century, neuroscientists investigated nonhuman primates (NHP) as a meso-scale model for understanding the anatomy, physiology and pathology of the human brain. Numerous NHP studies have provided valuable insights into neuroanatomy, function, development and aging, as well as providing a platform for understanding human brain disorders such as Parkinson's disease and mood disorders (see (Buffalo et al., 2019) for reviews). However, limitations in spatial fidelity and quantitative comparison between human and NHP brains have hampered accurate extrapolation across primate species. These limitations largely stem from dramatic differences in brain size, as well as principles of anatomical and functional organization, particularly in the cerebral cortex (Van Essen et al., 2019). Much of our understanding of cerebral cortex is centered around the concept of the cortical areas, as defined by architecture, function, connectivity, and topographic organization (Van Essen and Glasser, 2018). Accurate delineation of areas in a given species is extremely important for characterization of functions, but has proven to be extremely challenging. The macaque monkey has been one of the most intensively studied laboratory animals, becoming better understood in the latter part of the 20$^{th}$ century than other laboratory animals (and also humans), despite the many competing macaque parcellation schemes that were developed. In the ensuing decades, progress in parcellating mouse (Gămănuţ et al., 2018; Harris et al., 2019) and human cortex (Glasser et al., 2016a) has arguably advanced beyond that for the macaque (and perhaps the New World marmoset and owl monkey). It is noteworthy that state-of-the art human cortical parcellation combines 'non-invasive' and 'multi-modal' imaging of the whole brain including imaging-based estimates of interareal connectivity.

Species differences in cortical organization between human and macaque reflect evolutionary divergence since the last common ancestor some 25 million years ago (Kumar and Hedges, 1998). At the meso-scale, the divergence is greatest for association cortices - especially prefrontal, parietal, and lateral temporal regions engaged in higher functions such as cognition, affect and social behaviors including language. The disproportionate expansion of the human association cortex relative to the macaque (Van Essen and Dierker, 2007) is likely more modest relative to chimpanzees, which diverged 5-7 million years ago (Kumar et al., 2005). However, the definition of association cortex itself has not always been consistent, exemplified by disagreements as to how the extent of prefrontal cortex scales with overall brain size in primates (Barton and Venditti, 2013; Donahue et al., 2018; Gabi et al., 2016; Passingham and Smaers, 2014; Semendeferi et al., 2002). Patterns of brain connectivity have also changed over the course of evolution, such as the increased size and strength of the arcuate fasciculus (as revealed by diffusion MRI) arising together with the evolution of human-specific language functions (Rilling et al., 2012, 2008). Non-invasive neuroimaging is well positioned for objective and meso-scale comparison across species, since the same imaging or measurement methods can be applied in multiple species of interest.

Progress in human neuroimaging data acquisition and analysis include advances associated with large scale studies such as the Human Connectome Project (HCP) (Glasser et al., 2016b; D. C. Van Essen et al., 2012c) and the UK Biobank (Miller et al., 2016). Translating such advances to the arena of NHP neuroimaging has been slow, in part because conventional NHP MRI data acquisition has not been adapted to the specific challenges posed by NHP brains. Many researchers who study macaques use MRI scanners and coils built for humans, and some use ultra high-field scanners that often lack the pulse sequences support for parallel imaging. Non-invasive NHP imaging typically involves a small number of animals in each study. Importantly, these limitations are being addressed by recent technical advances such as high-quality head coils adapted to specific NHP brains (Autio et al., 2020b), cutting edge HCP-Style imaging acquisition protocols/preprocessing (Autio et al., 2020b), sharing NHP neuroimage data across sites (Milham et al., 2018) and ultra-high field neuroimaging (Ghahremani et al., 2016; Hori et al., 2020; Liu et al., 2020, 2019) (Yacoub et al., 2020). This enhances the prospects for validation studies combining non-invasive multi-modal imaging methods and invasive anatomical and electrophysiological methods that can best be carried out in NHPs.

Tract tracing shows that connections between cortical areas can be defined as feedback and feedforward (Rockland and Pandya, 1979), which leads to a hierarchical organization of the cortex (Felleman and Van Essen, 1991). Modern theories of the computational principles of cortical function are centered around understanding hierarchical processing of information (Friston, 2010; Markov and Kennedy, 2013; Mumford, 1992; Rao and Ballard, 1999; Vezoli et al., 2020). Cortical connections are defined as feedforward and feedback according to the laminar location of the cell body and the laminar target of the inter-areal axons using invasive anatomical tract tracing. These features cannot be detected with non-invasive imaging techniques and are exclusively available with tract tracing. More recently using principles drawn from structural hierarchy, there has been successful functional characterization of feedback and feed forward connections using electrophysiology in macaque and human (Bastos et al., 2015; Michalareas et al., 2016) allowing functional hierarchies to be defined in these species. In humans, laminar analyses using ultra-high resolution fMRI give insights into some aspects of hierarchy(Bergmann et al., 2019; De Martino et al., 2015; Kok et al., 2016; Muckli et al., 2015). The above considerations indicate that a deeper understanding of hierarchical processing in the cerebral cortex will require an accurate alignment of tract tracing and neuroimaging results, as we propose here.

Here, we describe our collaborative nonhuman primate neuroimaging and neuroanatomy project (NHP_NNP). Building on multiple pairwise collaborations between our laboratories, the NHP_NNP was launched in 2018 as an international collaboration across five laboratories (RIKEN, KU, WUSTL, SBRI/IoN, OU); UMinn joined in 2020. The project's goals include (i) implementing a high resolution HCP-style data acquisition and analysis approach in 100 macaques and marmosets; (ii) developing a comprehensive quantitative invasive tract tracing database accurately mapped onto MRI-based volumes and surfaces; (iii) comparing imaging-based functional and structural connectivity measures to "ground-truth" anatomical tract tracers; (iv) analyzing cytoarchitecture quantitatively via immunohistologically-defined cell types to map brain areas; (v) studying over 100 macaques and marmosets, to compare individual variability in social behavior, genetics and brain organization; and (vi) studying some macaque animals using a unique ultra-high resolution 10.5T scanner at UMinn. Achieving these objectives will entail validation of non-invasive multi-modal imaging methods and establishing combined neuroimaging and neuroanatomy atlases. Improved methods may also be translated to some extent to recent population neuroscience in humans, which collects 'big' data (~3K subjects in HCP projects, ~100K in UK Biobank) and will improve the quality of species-specific group average brain templates and their subject-wise variability. This perspective article summarizes our current activities and future project plans, which span neuroimaging, neuroanatomy, modeling, and combined approaches that exploit emerging technologies.

## 1.1. Overview of NHP_NNP

We are currently collecting *in vivo* neuroimaging data in large samples of macaques and marmosets, and we plan to acquire both neuroimaging and anatomical data from a smaller number of macaques. The *in vivo* data, including behavioral, genetics and high-quality brain imaging, are mainly acquired at the labs of Takuya Hayashi and Chihiro Yokoyama in RIKEN. The analysis of in-vivo brain imaging is carried out by Hayashi's lab and David Van Essen's lab at WUSTL. Anatomy studies are carried out in Henry Kennedy's lab (in SBRI, Lyon and IoN, Shanghai) with David Van Essen's lab at WUSTL contributing to data analysis. Some animals in Lyon and Shanghai are also scanned with high-quality brain imaging by the same protocol as in RIKEN. Stephen Smith's lab at OU is engaged in developing a core part of in-vivo image preprocessing and higher-level modeling of connectivity using in-vivo brain imaging data. David Van Essen's lab at WUSTL is involved in establishing cortical surface mapping of multi-modal data, surface-based analysis, and registration across species. The CMRR at UMinn (Director, Kamil Ugurbil) and the labs of Jan Zimmermann, Sarah Heilbronner, Damien Fair and Essa Yacoub carry out NHP imaging on a 10.5T scanner (Yacoub et al., 2020, this issue). The animal experiments in IoN, RIKEN, KU SBRI, and UMinn were approved by the relevant local animal experimental committee.

The project aims to obtain high quality *in vivo* multimodal data (i.e. brain MRI, social behavior, and genetics) from three genus (and five species): ~100 macaques (rhesus, *Macaca mulatta*; crab-eating monkeys, *Macaca fascicularis*; and Japanese rhesus, *Macaca fuscata*), ~100 common marmosets (*Callithrix jacchus*) and ~10 owl monkeys (*Aotus lemurinus*). All data except for those used in anatomical studies are being acquired in living animals. This is large compared to most published NHP imaging studies, which typically face practical limitations related to ethical and economic reasons, but is modest relative to large-scale human neuroimaging endeavors such as the HCP and UK Biobank. Individual variability is much lower in monkeys compared to humans (see Fig. 2), which in some respects alleviates the need for large sample sizes. At RIKEN, our strategy for macaques is to obtain data while macaques are hosted for a short period at our imaging facilities at RIKEN by generous arrangements with animal vendors. We plan to obtain data from adult macaques aged 3 years or older, sampling both sexes equally. For neuroanatomy, the project aims to investigate connectivity of a total of 139 cortical areas in each hemisphere by injecting retrograde tracers. We already have data from 52 cortical injections (51 neocortical injections + hippocampus); thus additional data will be obtained for ~ 90 injections. The tract tracing experiments are in a large part carried out in female cynomolgus monkeys. For marmosets, RIKEN Kobe has a breeding colony of over 120 marmosets, and we aim to obtain data from all animals in this colony once they reach maturity (age 2 years or older). In addition, we also study 10 owl monkeys generously provided by Prof. Masahiko Takada, Kyoto University Primate Research Institute.

## 2. Neuroimaging

### 2.1. Harmonized brain MRI protocols and preprocessing for comparative neuroanatomy

An important consideration when comparing brain organization across species is standardizing data acquisition and analysis protocols. Across the primate species under consideration, there is a ~200-fold difference in brain volume (~5-fold difference in isometric scale [cubic root of brain volume]), ~100-fold difference in cortical surface area (~10-fold difference in isometric scale [square root of surface]), and ~1.7-fold difference in median cortical thickness (Table 1). There are also major differences in modern estimates of the number of cortical areas in human (~180) (Glasser et al., 2016a; Gordon et al., 2016; Nieuwenhuys et al., 2015), macaque (91-161) (Markov et al., 2014a; Paxinos et al., 1999; D.C. Van Essen et al., 2012a) and marmoset (116) (Hashikawa et al., 2015; C. Liu et al., 2018; Majka et al., 2016; Paxinos et al., 2011) (see also (Van Essen et al., 2019) for review).

These species differences in size necessitate image acquisition protocols appropriately scaled for each species. In principle, one way to standardize data acquisition is to acquire a similar number of data points across species by adjusting the imaging resolution to the size of the brain or cortical surface area. However, simply scaling by a 200-fold range in brain volume equates to a 200-fold SNR penalty for the smallest voxels, which is not practically tolerable given currently available *in vivo* imaging systems. Instead, our approach is to meet a key minimal requirement by adjusting voxel resolution according to the thinnest parts of the cerebral cortex. This allows mitigation of partial volume effects across tissue categories (grey matter, white matter, and CSF) and distinguishing between opposing banks of sulci (Glasser et al., 2013).This results in practically achievable spatial resolutions (see Neuroimaging numbers in Table 1) and harmonizes the 1.7-fold difference in the cortical thickness across species (Fig. 1). Nonetheless, the spatial resolution scaled to the cortical thickness requires significant SNR boost for NHP neuroimaging. To obtain the needed SNR gains we designed and constructed species-specific multi-channel receiver coils for macaque and marmoset for use in a connectomics-optimized 3T scanner (MAGNETOM Prisma) (Autio et al., 2020b; Hori et al., 2018). These strategies enable us to utilize in the NHP the same multi-band accelerated EPI sequences as in the YA-HCP at much higher resolution for these smaller brains (see Fig 3. in Autio et al. 2020b).

**Table 1. Neuroanatomical and neuroimaging numbers in humans and NHPs.**

| NEUROANATOMICAL NUMBERS | Human | Chimpanzee | Macaque | Marmoset |
|---|---|---|---|---|
| **Brain[1]** | | | | |
| Brain size [$cm^3$] (mean ± sd) | 1395 ± 142 | 404 ± 32 | 90 ± 17 | 7.5 ± 0.5 |
| Volume ratio [%] | 100 | 29 | 6.5 | 0.53 |
| Isometric scale ratio of brain volume [%] | 100 | 66 | 40 | 18 |
| **Cerebral cortex[1]** | | | | |
| Surface area [$cm^2$/hemisphere] | 945 | 317 | 106 | 9.9 |
| Isometric scale ratio of cortical surface [%] | 100 | 58 | 34 | 10 |
| Cortical thickness (median) [mm] | 2.7 | 2.5 | 2.1 | 1.6 |
| Cortical thickness (5 percentile) [mm] | 2.0 | 1.7 | 1.4 | 0.9 |
| Cortical thickness (minimum) [mm] | 1.5 | 1.1 | 0.9 | 0.6 |
| **Number of cortical areas** | 178[2] 180[3] 200[4] | N.A. | 130[5] 161[6] 91[7] | 116[8,9,10] 54,106[11] |
| **NEUROIMAGING NUMBERS** | | | | |
| **Protocol[13]** | YA-HCP | Yerkes | NHP_NNP | NHP_NNP |
| **Spatial resolution of MRI [mm]** | | | | |
| Structural MRI | 0.8 | 0.8 | 0.5 | 0.36 |
| Functional MRI | 2.0 | NA | 1.25 | 1.0 |
| Diffusion MRI | 1.25 | 1.8 | 0.9 | 0.8 |
| **Temporal resolution of fMRI** | | | | |
| Repetition time (TR) [sec] | 0.72 | NA | 0.755 | 0.76 |
| #Time points per subject | 4800 | NA | 8092 | 8092 |
| Multi-band factor (cross-plane acceleration) | 8 | NA | 5 | 2 |
| **Diffusion weighting gradients** | | | | |
| b-values | 0,1000,2000,3000 | 0,1000 | 0,1000,2000,3000 | 0,1000,2000,3000 |
| # direction of diffusion encoding gradients | 270 | 60 | 500 | 500 |

1) Statistics of brain and cerebral cortex are from Human Connectome Project (YA-HCP) (N=1092 subject and 2184 hemispheres), YerkesChimp29 (N=29 subjects, 58 hemispheres) (Donahue et al., 2018), initial NHP_NNP macaque data, Mac30BS (N=30 subjects, 60 hemispheres) and NHP_NNP marmoset data (N=50 subjects, 100 hemispheres). The brain size and cortical area were estimated by brainmask_fs.nii.gz and midthickness surface in native T1w space, both calculated by HCP and HCP-NHP pipelines. 2) Gordon et al., 2016; 3) Glasser et al., 2016; 4) Nieuwenhuys, et al., 2015; 5) Van Essen et al., 2012a; 6) Paxinos et al., 2000; 7) Markov et al., 2014a; 8) Paxino et al., 2015, 9) Majka et al., 2016, 10) Hashikawa et al., 2015, 11) Liu et al. 2018. 12) Neuroimaging protocols are from the young-adult Human Connectome Project (YA-HCP) (Van Essen et al., 2012c), Yerkes for chimpanzee study (Donahue et al., 2018) and NHP Neuroimaging & Neuroanatomy Project (NHP_NNP) for macaque and marmosets (Autio et al., 2020b). The imaging protocols of NHP_NNP are available at https://brainminds-beyond.riken.jp/hcp-nhp-protocol/

    Structural images (both T1w and T2w) provide the foundation for accurate estimates of white matter and pial surfaces. The HCP's recommendation for acquisition of structural images at an isotropic resolution at least as small as half of the minimum thickness of the cerebral cortex (Glasser et al., 2016b) has been successfully adapted to the NHP brains in macaque (0.5 mm) (Table 1). Marmoset structural imaging resolution was set to the minimum allowed by the product sequence (=0.36 mm). Subsequent data acquisitions and analyses revealed a lower value for minimum cortical thickness (≈0.6 mm, Table 1 and Fig. 1). The T2w images allow more accurate pial surface estimation by reducing errors from dura matter and blood vessels, which are confounds in the T1w images, and to use the T1w/T2w ratio as a measure of cortical myelin (Glasser et al., 2014; Glasser and Van Essen, 2011). Fat suppression was applied to T1w scans to attenuate signals from bone marrow and scalp fat, thereby reducing potential confounds in estimation of cortical outer surface contours. Since movement during structural image acquisition can cause blurring and imaging artifacts, we acquire all structural (T1w and T2w) and diffusion images under deep anesthesia (see section 2.5).

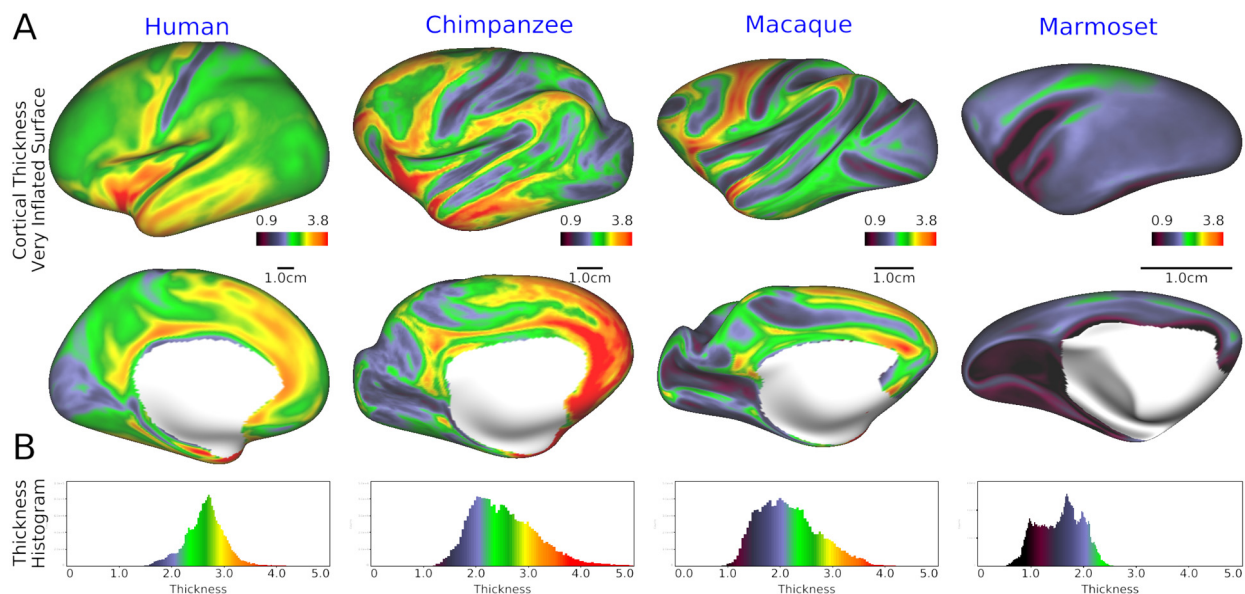

**Figure 1 Cortical thickness in human, chimpanzee, macaque and marmoset**
**(A)** Cortical thickness overlaid on a very-inflated left cortical surface (upper, lateral view; lower, medial view) in human, chimpanzee, macaque, and marmoset. **(B)** histograms of cortical thickness. The cortical thickness maps were created by averaging a population for each species (N=1092 in humans, 29 in chimp, 30 in macaque and 50 in marmoset). The median, lower 5th percentile and minimum of the averaged cortical thickness is 2.7, 2.0 and 1.5 mm in humans, 2.5, 1.7 and 1.1 mm in chimps, 2.4, 1.4 and 0.9 mm in macaque, and 1.6, 0.9 and 0.6 mm in marmoset (see Table 1). Data at https://balsa.wustl.edu/G39v3

    Data preprocessing used an HCP-NHP pipeline (Autio et al., 2020b; Donahue et al., 2016), adapted to each NHP species by modifying many parts of the HCP pipeline (Glasser et al., 2013). The major modifications are listed in Table 2 including: 1) measurement-related features and 2) neurobiological differences across primate brains. Measurement-related features include variations in $B_0$ and $B_1$. Although RF transmission in clinical 3T scanners yield relatively homogeneous $B_1$ over the small imaging volumes needed to image NHP brains, the actual $B_1$ field suffered from non-negligible degradation over the macaque brain by several degrees (in flip angle) that is thought to be due to interference with the multi-channel RF receiver (see Fig. S3 in Autio et al. 2020b). Therefore, additional bias field corrections were incorporated to the T1w and T2w images prior to the automated segmentation procedure used for cortical surface reconstruction. Cortical myelin maps, calculated from T1w and T2w ratio, can be corrected for bias with a reference myelin map for neuroanatomical purposes (Glasser et al 2013); however, further refinements are expected in the future for residual transmit field (B1+) related biases.

    Many of the adaptations of HCP-style preprocessing involve adjustments related to NHP differences in brain size relative to humans (Table 2). Differences in brain size are expressed by an isometric scale factor, which is approximately 0.6 in chimp, 0.4 in macaque, and 0.2 in marmoset relative to that in humans (Table 1). Accordingly, various brain size-related parameters are scaled down by these factors during preprocessing (i.e. field of view of 3D volume atlas space, and warp resolution of atlas registration and EPI distortion correction). In the FreeSurferPipeline, the original brain image is scaled up with this ratio to get its size close to the human's brain, so that FreeSurfer's surface reconstruction works reasonably well, and the outputs of FreeSurfer are rescaled back to the original brain size. A FreeSurfer Gaussian Classifier Atlas (GCA), a probabilistic structure map, was created based on manually classified 21 structures using a training set of 21 subjects (Autio et al., 2020b) and used for automated subcortical segmentation in each species. The GCA can identify the less myelinated claustrum, which is needed for precise white matter surface estimation beneath insular cortex (see example in Fig. S5 of Autio et al. 2020b) thereby avoiding a 'claustrum invagination problem' in Fig. S6 of Autio et al., (2020b). A species-specific cortical thickness maximum is used to prevent underestimation of cortical thickness in these upscaled datasets. The less myelinated and thinner white matter blades in anterior temporal and occipital cortex in NHP are addressed by feeding a white matter skeleton into the FreeSurferPipeline (Autio et al. 2020b). Less cortical folding in smaller NHPs such as marmosets requires adjustment of the initial rotation in the

FreeSurfer surface registration (Fischl et al., 1999) and a surface registration template based on folding is created for each species. In humans, HCP-style surface registration uses the affine alignment of folding maps provided by FreeSurfer (Fischl et al., 1999), followed by symmetrization of the left-right hemisphere (Van Essen et al., 2012b) and finally with a gentle nonlinear alignment using folding maps (MSMSulc) (Robinson et al., 2014) and multi-modal metrics (MSMAll) (Glasser et al., 2014; Robinson et al., 2018). In NHPs, MSMSulc registrations are implemented across species, where the MSMSulc in NHPs works well and results in small distortions within a neurobiologically-expected range (Robinson et al., 2018; Van Essen, 2005). It is not yet known whether NHPs will benefit from a multi-modal registration like MSMAll (using modalities such as myelin maps, resting-state networks, and visuotopic maps), given their reduced individual variability.

**Table 2. Modifications of HCP-style preprocessing for NHP**

| Category of modification | Class | Modified or additional items in preprocessing pipeline |
|---|---|---|
| **Measurement features** | Static magnetic field ($B_0$) | Optimized distortion correction using $B_0$ fieldmap and TOPUP, Note that this depends on both the resolution of imaging and the size of objects (e.g. brain) (see below). |
| | Radiofrequency field ($B_1$) | Optimized $B_1$ biasfield correction in FreeSurferPipeline (IntensityCor). The $B_1$ bias field depends on the MR system (e.g. strength of magnetic field, configuration of transmitter and receiving coils) and objects (e.g. size and electromagnetic properties) |
| | Number of Wishart distributions | Computation of variance (VN) and estimation of dimensionality of fMRI data in ICA (icaDIM), which may depend on the interaction of the random noise of the measurement system with the voxel size, brain size, and processing steps |
| | Step length of diffusion tractography | 25% of sampling resolution, depending on the spatial resolution of diffusion MRI |
| **Neurobiological features** | Field Of View (FOV) | Brain templates for T1w, T2w and subcortical ROIs (Atlas_ROIs), isometrically scaled for brain size |
| | Non-linear warp resolution | Configuration in FNIRT and TOPUP, isometrically scaled for brain size |
| | Scaling for FreeSurfer | Scaling and rescaling of brain image in FreeSurferPipeline, isometrically scaled for brain size |
| | Cortical thickness | Maximum cortical thickness in FreeSurferPipeline, adapted for histogram of cortical thickness in each species |
| | Cortical gyrification | Optimized max angle for search in surface registration (Marmoset = 50, Macaque = 68, Human = 68), which depends on the complexity of gyrification |
| | | Template of surface registration in FreeSurfer (i.e. $GCAdir/?h.average.curvature.filled.buckner40.tif) and reference sulc for MSMSulc. Created for each species and used as a template of surface registration. See also effects of surface registration in Figure 1S. |
| | Brain atlas and segmentation | Brain templates for structure MRI (T1w, T2w) and Atlas_ROIs. Species specific templates. Skull stripping of structure MRI of NHPs is also useful for initialization of registration |
| | | Brain segmentation atlas for each species (i.e. Gaussian Classifier Atlas (GCA) at $GCAdir"/RB_all_2008-03-26.gca aseg.auto_noCCseg.mgz) |
| | | White matter skeleton feeding in FreeSurferPipeline, adapted to thin white matter blades in NHPs. Needed for correct estimation of white surface. |
| | Registration of brain in fMRI | Use mcflirt_acc.sh for motion correction, use SE-EPI and T2w for initialization, use T1w-contrast FreeSurfer BBR for MION fMRI |
| | Artifact removal in fMRI | ICA+FIX training file, specific to species's brain, measurement systems and contrasts (BOLD vs MION) |
| | | Standard venous sinus maps in NIFTI volumes used for ICA+FIX |
| | Myelination | Population-based bias field correction of cortical myelin map, plus subject or scan-wise transmit field bias field correction is planned in future. |

See also some of these features in the flow chart of the HCP-NHP pipelines in Fig. S1 in (Autio et al 2020b).

## 2.2. Surface-based analysis for NHPs and comparison with human

Across primates, cortical organization exhibits many similarities but also distinct differences in size and variability in structural and functional organization of the cerebral cortex (see Fig. 2 for the left hemisphere). For example, among primates, cortical folding is most pronounced in humans, followed by apes, including chimpanzees (Fig. 2A). The average midthickness cortical surface after nonlinear surface-based registration across subjects is shown in Fig. 2B. Note that in humans the average cortical midthickness surface does not preserve all of the individuals' cortical folding patterns (due to greater between-subject variability than seen in NHPs) but rather exhibits a smoother, less folded cortical surface. By comparison, in chimpanzees, the population average is more similar to the individual pattern of convolutions, and this is even more pronounced in the macaque. In lower NHPs, particularly in marmoset, both the individual and population average cortical midthickness surfaces are modestly gyrified (i.e. smooth) and remarkably similar in topology. The 3D variability in cortical midthickness surface (Fig. 2C), adjusted according to the isometric scale of the brain, indicates that the smoothness of population-averaged cortical surface in humans is due to the large cross subject variation in the cortical folding patterns, and considerably higher variability in 3D midthickness is found particularly in association areas in humans relative to NHPs. Chimpanzees also exhibit noteworthy variation in association areas but these subject-variations become progressively smaller in macaque and marmoset monkeys. These findings highlight the dramatic inter-subject variability of 3D cortical architecture (gyrification, sulcal formation) in higher primates' association areas, explaining why 3D volume-based approaches traditionally performed in human neuroimaging lead to severe blurring and loss of accurate neuroanatomical localization (Coalson et al., 2018).

Population average cortical myelin maps were generated by mapping the T1w/T2w ratio to each individual midthickness surface, surface-registered using MSMAll (human) or MSMSulc, and corrected for bias fields (Glasser and Van Essen, 2011; Robinson et al., 2018). This reveals the relatively higher myelination in the primary sensorimotor, auditory, visual and MT areas compared to association areas in humans (Fig. 2D). This pattern is comparable between humans and NHPs including marmosets (Van Essen et al., 2019), although the relative size of the lightly myelinated association cortex (blue to black area) becomes progressively larger going from smaller to larger primates (Glasser et al., 2014). In humans and chimpanzees, despite high inter-subject variability of 3D folding patterns in association areas, myelination in these areas is less and relatively constant across subjects. These findings demonstrate the usefulness of surface-based approaches to minimize subject variability and increase species-specificity in higher-order cognitive areas of higher primates, and raise interesting issues concerning subject variability in higher cognitive areas (Mueller et al., 2013).

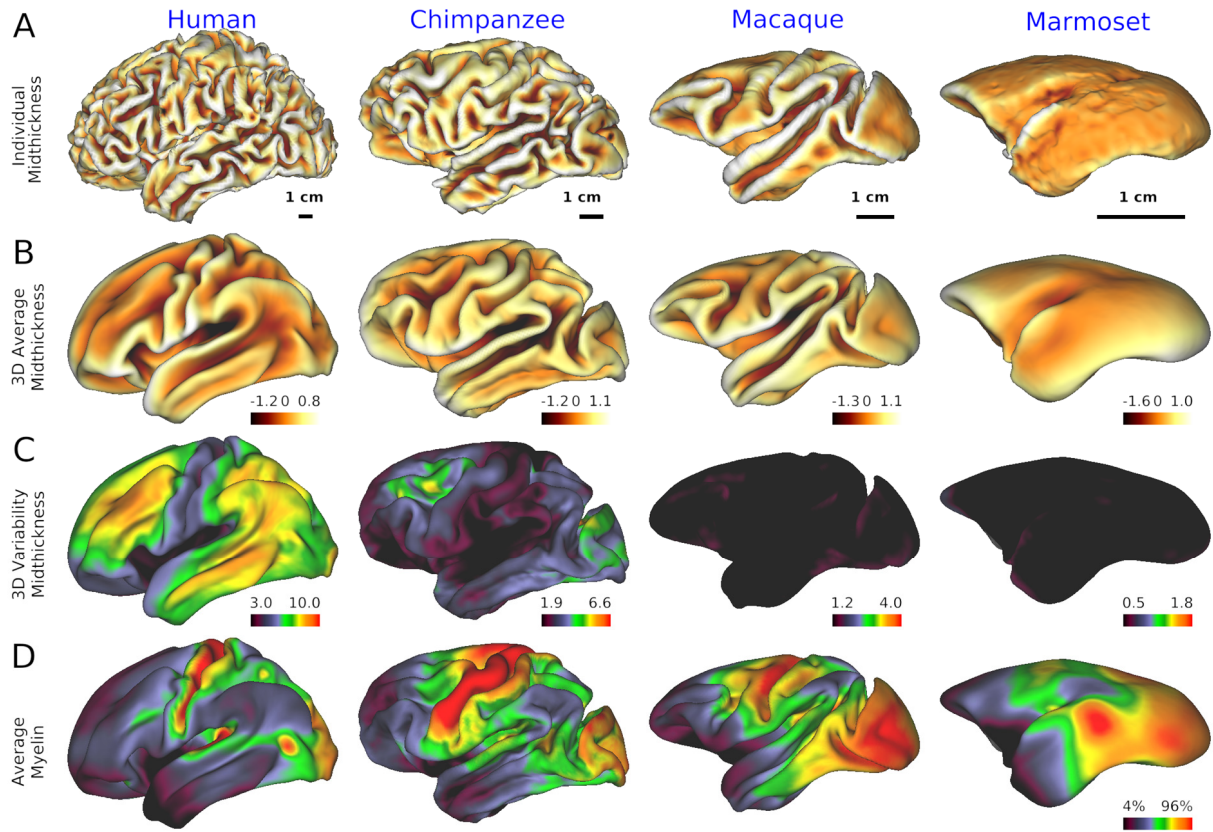

**Figure 2. Cross-species comparison of individual's cortical surface, average and variability**
**A**) cortical midthickness surface in 32k mesh in a single subject. At the single-subject level, cortical folding is more prominent in humans than in other species. The cortical surface is color coded by sulcal depth in the orange-yellow (see color bar and range in the right lower corner in each panel in second row) **B**) 3-dimensional (3D) average of midthickness surface in 32k mesh, Note that the average cortical surface does not follow the individual's cortical folding pattern but exhibits relatively smooth cortical surface area, whereas marmoset individual and average cortical surfaces are both smooth and remarkably similar. The 3D variability in cortical midthickness, colormap adjusted by isometric scale of the brain (see Table 1), suggests that this smoothness of the average cortical surface is due to the large cross subject variation in folding patterns. Cortical surface is color coded by sulcal depth. **C**) Variability (3D standard deviation) of midthickness surface with a colormap range scaled across species by isometric scale of brain size (see Table 1). Note that in humans and chimps, association areas exhibit high 3D variability of midthickness surface as compared with primary sensorimotor, visual, and auditory areas. **D**) average myelin contrast in human (N=1092), chimpanzee (N=29), macaque (N=30) and marmoset (N=50). Average cortical myelin contrast (estimated from the T1w/T2w ratio) is high in primary sensorimotor, visual and auditory and MT areas in all species but lower in the association areas. Data at
https://balsa.wustl.edu/L6xn7

Subcortical structures in the human brain are estimated to include 455 regions (Fcat, 1998) that differ in topology, architecture and shape. The shapes of various subcortical nuclei are complicated and vary from globular (e.g. thalamus, amygdala, pallidum), elongated (e.g. caudate) to sheet-like (e.g. claustrum). Therefore, subcortical structures are represented as voxels embedded in the species-specific 3D standard space (e.g., Montreal Neurological Institute [MNI] Space). However, most subcortical nuclei are too small to clearly visualize with the resolution of current non-invasive imaging techniques: only 7 % among 455 regions are visible in MRI in humans (Forstmann et al., 2017). The 3D registration of the subcortical structures would benefit from refinement, for example by using diffusion fiber orientation information (Zhang et al., 2006).

CIFTI (Connectivity Informatics Technology Initiative) greyordinates are a combination of cortical surface vertices on a standard mesh and subcortical voxels in a standard volume space (Glasser et al., 2013). This format is useful not only for standardizing across subjects but also across species. For example, cortical coordinates are not linked to an exact location in 3D space, but rather to the vertex on a particular 2D surface mesh. The 2D surface mesh can be inflated or made spherical and can be registered across subjects or even across species. Hence, data in the CIFTI format is well positioned to enable cross-species cortical comparison, even when there are large differences in brain size. Figure 2 shows all the cortical maps in 32k vertices (in one hemisphere) across four species and overlaid on a 'surface model' of midthickness (however, evolutionarily corresponding regions have not been brought into alignment in the surfaces of different species - see Section 2.6.) The same surface mesh is useful to compare the distribution of metrics over cortex across species, while species-specific meshes can be used to approximate the same neuroanatomical spacing of surface vertices and resolution for the data acquisition (Table 1). For subcortical structures, voxels are parcellated by FreeSurfer into a total of 21 regions (10 regions in left and right, plus brainstem) aligned across subjects in standardized 3D space and embedded into the species-specific CIFTI greyordinates. The CIFTI format is useful not only for MRI data, but also for any modalities such as histology and gene expression that can be represented on 2D cortical surface vertices and 3D volume voxels. Therefore, multi-modal data can be simultaneously handled and directly compared using identical indices, which is particularly advantageous for establishing future combined atlases for neuroimaging and neuroanatomy atlases (see section 4.2).

## 2.3. Functional MRI in NHP

Our fMRI strategy at RIKEN includes two approaches: i) study a relatively large cohort of animals (N≈100) in a lightly sedated condition during resting state fMRI BOLD scans with a duration of 102min x 2 scans per subject, and ii) in a relatively smaller number of animals (~10) obtain a large number of repeated scans in the awake and anesthetized conditions and using monocrystalline iron oxide nanoparticles (MION, Feraheme®, AMAG pharmaceuticals Inc, MA) to enhance contrast-to-noise (Vanduffel et al., 2001). The first approach allows exploration of species-specific network functional organization, intersubject variability, and the links between genetics and behaviours (see section 6). The second approach aims to provide the high contrast fMRI that is necessary to reveal species-specific networks in awake conditions, thereby facilitating more accurate comparison of awake resting-state networks with those in humans.

Among the many different fMRI parameters, spatial and temporal resolution are particularly important. For planning spatial resolution, we applied the neuroanatomically informed approach based on cortical thickness (see Table 1). In humans, a voxel size of 2mm was used, which is at the 5th percentile of the human cortical thickness histogram (Glasser et al., 2016b, 2013) and minimizes the partial volume effect in the direction perpendicular to the cortical surface. This approach was applied to NHPs and set the fMRI spatial resolution of macaques and marmosets to 1.25 mm and 1.00 mm, respectively (Table 1). We applied a neurophysiologically informed approach to set the temporal resolution of functional MRI. Resting-state functional networks have the strongest power in the low frequency range of 0.01 to 0.5 Hz in humans, and this is expected to be similar in nonhuman primates (Buzsáki et al., 2013). Additionally, artifactual or undesired signal from head motion, physiology, and the scanner often has differing and higher temporal frequencies. Therefore, the target temporal resolution of the fMRI scanning in NHP was TR≈0.7 sec as in human studies (Table 1).

As described earlier, we adopted custom multi-array coils for NHPs to achieve the desired high spatial and temporal resolution fMRI in NHPs. The target resolutions (Table 1) are achieved by increasing the number of coils to 24 in macaque (Autio et al. 2020b) and 16 in marmoset, combined with parallel imaging sequences (Feinberg et al., 2010; Moeller et al., 2010; Setsompop et al., 2012; Wiggins et al., 2006; Xu et al., 2013). The temporal signal-to-noise ratio (tSNR), calculated by the mean divided by standard deviation of preprocessed uncleaned fMRI data (Smith et al., 2013), reveals reasonably high values in grey matter of macaque (53±27) (Autio et al., 2020b) and marmoset (~100), which are compatible with those in human from YA-HCP (~40) (Smith et al., 2013). This tSNR gain in NHP is primarily attributable to closer proximity of the RF receive coils and to the smaller diameter of the receive coil elements (human≈80 mm, macaque≈50 mm and marmoset≈20 mm) and to overall head coil diameter (human≈200 mm, macaque≈100mm, marmoset≈50mm).

Another important factor for reliable fMRI is the scan duration (Birn et al., 2013; Laumann et al., 2015). In anesthetized animals, we achieved a total scan duration (102 min, see Table 1) within the available window of steady anesthesia and physiological conditions (see section 2.5 for anesthetic protocols). We confirmed that our macaque protocol provides higher sensitivity of functional connectivity (e.g. default-mode network) even in 15 min duration as compared to other PRIME-DE open data, and increasing the scan duration to a total of 102 min further enhances the quality of the fMRI as shown in [Fig. S12](#) in Autio et al., (2020b). The MION protocol in awake animals likewise involves multiple fMRI runs performed across days to maximize the total amount of fMRI data. Overall we plan to acquire around 1 million resting state fMRI volumes in the macaque, 1/5th of the 5 million resting state fMRI volumes acquired in the YA-HCP.

The preprocessing of resting-state fMRI data uses a cortical-surface approach to maintain the spatial fidelity of functional images (Autio et al., 2020b). Motion correction uses the mcflirt_acc.sh script in the HCP-NHP pipeline, which outperforms registration to reference volume. This showed robust results compared with mcflirt that quickly calculates 4D volume registrations but often results in motion estimation errors particularly in the smooth-shaped marmoset brain or low SNR data like MION fMRI. The fMRI data is corrected for distortion with spin-echo (SE) EPI and TOPUP using an optimized warp resolution (Table 2). SE-EPI data is used so that the registration-based internal correction model does not have to consider areas of differential signal dropout in the gradient echo images. The two opposing phase SE-EPIs will be obtained, which should be distorted locally by the same amount but in the opposite direction along the encoding axis. In TOPUP, the two volumes are non-linearly and symmetrically registered to meet halfway, so that the warp field is calculated to generate distortion-corrected SE-EPI images (Andersson et al., 2003). The non-distortion corrected SE-EPI is registered to the distorted fMRI, since we assume the same distortion between SE-EPI and GE-EPI by obtaining them with the same settings in phase encoding direction and echo spacing, then the warp field will be applied to fMRI data for correcting distortion. The registration between EPI and structural images is initialized using distortion corrected SE-EPI to the T2-weighted volume, and further fine-tuned with a FreeSurfer boundary-based registration (BBR) (Greve and Fischl, 2009) to the T1w-weighted volume. The data was mapped into species-specific greyordinates. Artifacts and nuisance signals are removed using automated classification (FIX) of independent component analysis (ICA) (Glasser et al., 2018; Salimi-Khorshidi et al., 2014). FIX is trained for each species (macaque, marmoset) and protocol (BOLD or MION fMRI) and achieves high accuracy of classification (Autio et al., 2020b). A venous sinus map in the standard space was created for each species to be used as a feature for the FIX classifier (Table 2, Fig. 3).

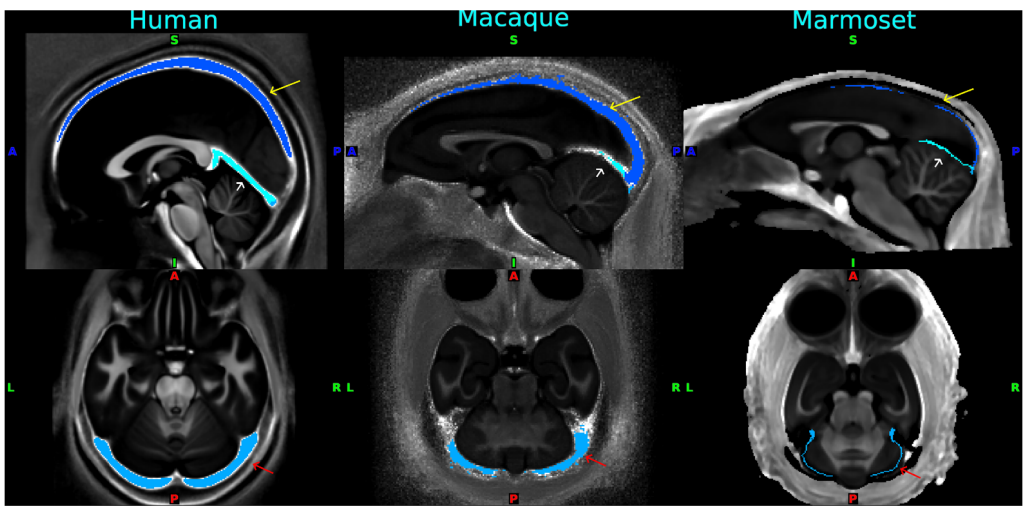

**Figure 3. Standardized venous sinus maps for each species of human, macaque and marmoset**
Yellow arrow: superior sagittal sinus, white arrow: straight sinus, and red arrow: transverse sinuses. A standardized venous map is used for creating a subject-specific venous map, which is then used for extracting the time series as a feature when automatically denoising functional MRI data using ICA+FIX. The venous maps were overlaid on the standardized map of T1w divided by the T2w image. Data at https://balsa.wustl.edu/pkqlD

Fig. 4 shows exemplar results for functional connectivity seeded from the posterior cingulate/precuneus cortex (pC/PCC), i.e. 'default mode network (DMN)' (Raichle et al., 2001; Vincent et al., 2007). An initial test fMRI dataset in anesthetized macaques in NHP_NNP (Mac30BS, N=30) was used for analysis of functional connectivity, while the human data was from young adult HCP subjects (N=210). The seed was placed in the left pC/PCC (shown by white sphere in Fig. 4). In both species, functional connectivity is distributed over widespread cortical regions involving parieto-temporal-frontal areas. Interestingly, functional connectivity suggests potential homologue regions in the lateral parietal cortex (PGi/PGs vs 7a/DP in human and macaque, respectively) and the dorsal prefrontal areas (8Av/8Ad/9p/8BL vs 8m/8B/9/46d/F7) (Fig. 4, bottom). These findings demonstrate that the cortical distribution of the 'default system' can be investigated with high spatial fidelity, which merits further investigations of the dynamics, variability and complexity of whole-brain resting-state networks across species. Accurate registration of cortical features across species to identify functional correspondences and evolutionary homologies is a topic of high importance (see section 2.6).

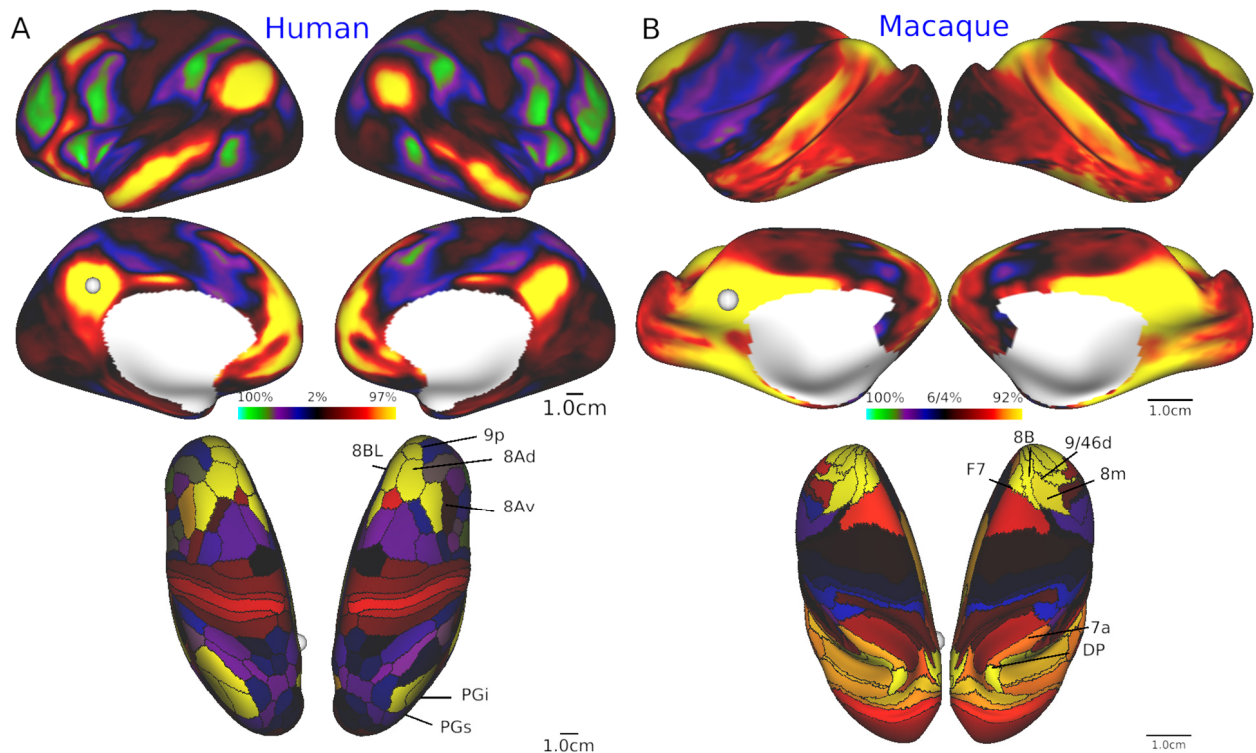

**Figure 4. Default mode network in human and macaque.**
Seed-based functional connectivity showing a typical cortical distribution of 'default mode network'. The seed (white circle) was placed in the left posterior cingulate/precuneus cortex (pC/PCC) of humans (HCP, N=210) and macaques (NHP_NNP Mac30BS, N=30). **A)** Dense functional connectivity maps (Upper row, lateral view; middle row, medial view) and the dense by parcellated functional connectivity maps (bottom, dorsal view) of the seed at pC/PCC in humans. **B)** Dense functional connectivity maps (upper row, lateral view; middle row, medial view) and the dense by parcellated functional connectivity maps (bottom, dorsal view) of the seed at pC/PCC in macaque. Functional connectivity is calculated with Pearson's correlation of the seed time series signal. The parcellation of the cerebral cortex is based on Glasser et al., 2016 in humans and on Markov et al., 2014a in macaques. Data at https://balsa.wustl.edu/97rzG and https://balsa.wustl.edu/kNq56

## 2.4. Behavioral state during functional neuroimaging - sedated, awake resting and task-performing

A critical issue for comparing brain function between species is the harmonization of 'brain state' during tests. What is the optimal 'state' for recording resting-state functional MRI in NHP and humans? Arousal state is known to affect the amount of global signal in the brain (Liu et al., 2017; X. Liu et al., 2018; Power et al., 2017; Tagliazucchi and Laufs, 2014; Yeo et al., 2015), though it is not yet well understood to what extent this effect is due to physiological confounds. In YA-HCP, resting-state fMRI data was acquired after subjects were instructed to remain awake, however, a substantial number of subjects fell asleep because of the relatively long scan time (15 min). For species comparison, it is ideal to obtain the NHP resting-state fMRI scans in the awake condition, however, this is practically challenging to achieve in a large cohort, since awake imaging requires an invasive surgery to implant a head post, and laborious training for animals to adapt to noisy fMRI environments. Therefore, we have evaluated various sedation protocols for resting-state networks in NHPs, leading to the decision to use a mixed sedation with low dose dexmedetomidine (4.5 mcg/kg/h) and isoflurane (0.6%) in NHP with mechanical respiration control using a gas mixture of air 0.75 L/min and oxygen 0.1 L/min. This protocol allows significant preservation of functional connectivity (see Fig. 4B and also Autio et al., 2020b). The detailed sedation protocols, procedures and animal management during NHP MRI scans are available at https://brainminds-beyond.riken.jp/hcp-nhp-protocol/.

The action of mixed sedatives on functional connectivity is largely unknown, but a mild sedative effect may be achieved by low dose dexmedetomidine. Dexmedetomidine is an agonist of α2-adrenergic receptors, has actions of sedative, anti-anxiety and pain relief and minimal impact on neuronal function and respiratory depression, and is commonly used for neurosurgery (Lin et al., 2019). In rodents, dexmedetomidine has been proposed as being suitable for resting-state fMRI studies (Pan et al., 2015; Bortel et al., 2020) by providing stable physiological conditions when given at a rate of 100 mcg/kg/hr (Sirmpilatze et al., 2019) or 75 mcg/kg/hr (Bortel et al., 2020). The dexmedetomidine dose in our NHP protocol was set to a lower level (4.5 mcg/kg/h) because of its cardiac suppression, but higher than that used in human anaesthesia (0.2-0.7 mcg/kg/hr). We included a small dose of isoflurane (0.6%) in order to maintain a stable anesthetic condition and to preserve functional connectivity and practical utility as compared with a mono-anesthetic protocol such as isoflurane 1.0% or intravenous propofol (5-10 mg/kg/hr), used in earlier studies (Hayashi et al., 2013, 2004; Kikuchi et al., 2017; Ohnishi et al., 2004). Combining dexmedetomidine and low-dose isoflurane is also recommended in recent functional studies in rodents (Bortel et al., 2020). Actions of the sedatives on brain state and functional connectivity needs to be further addressed in this project by scanning macaques with fMRI in both awake and anesthetized conditions using otherwise matched methodologies.

Task fMRI studies are also important in order to identify homologous functional landmarks across species. Previous studies revealed cross-species networks for oculomotor function (Hutchison et al., 2012), cognitive set shifting (Caspari et al., 2018; Nakahara et al., 2002), cognitive-motor behavior (Caminiti et al., 2015), and auditory function for harmonic tones (Norman-Haignere et al., 2019).

### 2.5. Diffusion MRI in NHP

High resolution diffusion MRI (dMRI) of NHPs in a clinical MRI scanner is more challenging than structural and functional MRI, due to limitations in gradient slew rate and strength. Ex vivo ultra-high field MRI studies are better positioned to provide more detailed description of NHP white matter neuroanatomy (Liu et al., 2020). Nonetheless, in vivo dMRI applications continue to have an important role in validating the imaging technologies and analysis protocols used for humans, and also enable controlled longitudinal investigations of plasticity of NHP white matter (Scholz et al., 2009; Sexton et al., 2014; Takenobu et al., 2014). Thus, our protocols in Takuya Hayashi's lab in RIKEN include in-vivo whole-brain diffusion MRI acquisition with isotropic resolution of 0.9 mm in macaques and 0.8 mm in marmosets and with a large number of diffusion gradient directions (500 directions) and b-values of 1000, 2000 and 3000 s/mm$^2$ (Table 1). The high resolution (0.3 mm) multi-shell ex-vivo dMRI of macaque brains has also been acquired in Henry Kennedy's lab in Lyon with b-value of 2000, 4000 and 6000 s/mm² and 64 gradient directions per shell.

The dMRI data analysis utilizes modified HCP-pipelines (with corrections for $B_0$ and eddy current distortions, and motion followed by fiber orientation estimation using 'bedpostx_gpu' and probabilistic tractography using FSL's 'probtrackx2_gpu' algorithms) (Autio et al., 2020b; Behrens et al., 2003; Donahue et al., 2016). Moreover, our dMRI analysis also contain species-optimized mapping of neurite orientation dispersion and density imaging (NODDI) which is used to evaluate tissue microstructure associated with neurite composition (a collective term referring to both dendrites and axons) (Fukutomi et al., 2018; Zhang et al., 2012).

### 2.6. Cross species registration

For quantitative comparisons across species, a critical issue is how best to align the cerebral cortex between humans and NHPs. There are large species differences not only in cortical surface area and complexity of convolutions (Fig. 2), but also in the layout of cortical areas and their relationship to cortical folds, particularly in association cortices. The smaller number of cortical areas in monkeys vs humans (see Table 1) suggests that NHPs likely lack a number of cortical areas that are present in humans, thus making precise area-to-area registration challenging. A number of cortical areas and regions can be proposed to be homologous (having a common evolutionary ancestor) based on a combination of features such as high myelin content in primary sensorimotor, auditory, and visual areas plus the MT+ complex (see Fig. 2). Early efforts used a limited number of cortical landmarks of presumed homologous areas and an inter-species registration algorithm having significant methodological limitations (Orban et al., 2004; Van Essen and Dierker, 2007). Several recent studies use discrete landmarks combined with feature maps, including myelin maps (Eichert et al., 2019), resting-state fMRI (Xu et al., 2019) or both myelin and fMRI as we have done (Donahue et al., in preparation). Our approach, a collaboration between the Van Essen and Hayashi labs, uses the chimpanzee Yerkes29 atlas surface (see Fig. 3) as a neutral 'middle ground' for bidirectional registration between human and chimpanzee and between macaque and chimpanzee. Each of these more recent approaches uses the MSM algorithm (Robinson et al., 2018) but in different ways, to constrain registration between human and macaque data mapped to spherical surfaces. Each achieves alignment that is likely to represent an improvement over the aforementioned older studies, but the recent registrations differ substantially from one another. It remains unclear which provides the best interspecies mapping in different cortical regions. Important issues requiring further development include accounting for the large distortions that occur when inflating anatomical (midthickness) surfaces to make the spheres used for registration, dealing with incompatibilities across species (including functional networks and cortical areas present in only one species), and performing registrations in a manner that facilitates hypotheses-driven evaluation of different combinations of multimodal data used to drive alignment.

## 3. Neuroanatomy

Anatomical studies of long-distance connections that link cortical areas in the macaque have provided major insights into organizational principles of the cortex and have helped shape theories of information processing in the brain. Early investigations of hierarchical organization and network properties of the cortex used data collated from numerous publications (Bakker et al., 2012; Felleman and Van Essen, 1991; Kötter, 2004). The interpretation of such studies was hampered by inconsistencies in the delineation of many cortical areas and by inadequate quantification of the strength and the laminar distribution of connections of each pathway. Major progress in addressing these issues has come from quantitative analyses of retrograde tracer injections into 29 cortical areas using a consistent 91-area atlas parcellation (Markov et al., 2011, 2014a). By systematically quantifying the strength of connections in terms of numbers of parent neurons in pathways linking areas, these and follow-up studies revealed an unexpectedly wide range of connection strengths (five orders of magnitude), a much higher matrix density (percentage of possible connections actually existing) and other important organizational features (Ercsey-Ravasz et al., 2013, p.; Knoblauch et al., 2016; Markov et al., 2013; Roberts et al., 2016). Our NHP_NNP collaborative team plans to extend this approach in several ways, including tracer injections into many more cortical areas, analysis of 'dense' (i.e. measured at the surface vertex-wise or voxel-wise level) connectivity patterns as well as 'parcellated' (i.e. brain area-wise) connectivity, and incorporation of more sophisticated information-theoretic approaches for modeling the parcellated connectivity graph.

### 3.1. A next-generation cortical connectivity map

*Increased spatial coverage.* Given the importance of extending connectivity analyses beyond the 29 cortical areas injected by Markov et al., (2014a), the Kennedy labs (Lyon, Shanghai) have continued making additional retrograde tracer injections. In addition to the traditional tracers DY (Diamidino Yellow) and FB (Fast Blue), Cholera Toxin B (CTB) linked to different-colored fluorophores is also used routinely. To date, multiple tracer injections have been made into 51 cortical areas (including the 29 already published) using an atlas of 91 areas. We plan to register existing injections to a new atlas of 139 areas developed in the Kennedy lab, which will include a finer parcellation of frontal areas. Data acquisition has been considerably accelerated using high-throughput digital slide scanning. We aim to inject tracers in the full complement of areas in the 139 areas, which will involve injecting an additional 90 areas. These injections will include difficult-to-access regions of temporal and orbitofrontal cortex, as well as the so-called rim areas along the medial wall. The objective is to complete the full complement of 139 areas by 2025. Injections of tracers are carried out in a stereotypical fashion with injections spanning 2.0 to 3.0mm at an angle of approximately 30 degrees so that the injection and the uptake zone is confined to layers 1 to 6 with minimal involvement of the superficial white matter.

The primary quantitative data provided by this approach is the precise location of each retrogradely labeled neuron in each of many closely spaced histological sections throughout the forebrain. For each section, digitized contour lines are drawn to delineate the pial surface, layer 4, the gray-white boundary, and (in many cases) the claustrum as well as numerous subcortical structures. For parcellated analyses, the number of labeled neurons in each area is normalized with respect to the total number of labeled neurons and is expressed as a FLNe (fraction of extrinsically labeled neurons) weight index. An associated index of laminar distribution is the fraction of Supragranular Labeled Neurons (SLN) within each area.

*Repeatability.* Earlier studies (Barone et al., 2000; Markov et al., 2011, 2014a, 2014b) exploited repeat injections into the same cortical area in different individuals, providing valuable estimates of the reproducibility of connection strengths and laminar distributions, showing that projections to areas V1, V2, V4, and 10 were generally consistent to within one or two orders of magnitude, which is far less than the total range of 5 orders of magnitude. In order to extend this approach to additional areas and to evaluate repeatability using a parcellation-free approach, we plan to repeat injections in the same location of a given cortical area for ~20 additional injections to quantitatively assess the reproducibility of the invasive connectivity maps and to validate our alignment procedures. The average correlation for parcellated and for dense (vertex-wise) connectivity across all repeat injection cases will set an upper bound for the maximum expected correlation of non-invasive and invasive connectivity (see below).

*Inter-hemispheric connectivity.* Previous studies indicate that the pattern of contralateral connectivity is weaker, but approximately symmetric to that seen on the side ipsilateral to an injection (e.g., (Lewis and Van Essen, 2000)), but a detailed quantitative comparison has been lacking. In all of our injection cases, we plan to map contralateral as well as ipsilateral tracer-based connectivity. We will then compute the correlation of injected side vs contralateral side connectivity for dense (vertex-wise) and parcellated maps, taking advantage of the left-right 'geographic' correspondence of the macaque Yerkes19 surface mesh (Donahue et al., 2016; Van Essen et al., 2012b). This will provide a critical baseline for analyses of interhemispheric connectivity patterns estimated using non-invasive functional connectivity and structural connectivity methods.

*Mapping 'dense' connectivity to a surface-based atlas.* The Van Essen and Kennedy labs have developed a 'parcellation-free' mapping of tracers to a cortical surface-based atlas (OHBM 2019 abstracts; Hou et al., in prep). This 'contours-to-atlas' surface mapping of tract tracing data serves three broad aims: firstly, to enable high-resolution analyses that preserve fine-grained detail on data mapped to an atlas surface, allowing comparisons with other tracer injection results as well as data from other modalities (e.g., cortical myelin maps, functional connectivity from fMRI, structural connectivity from dMRI, task fMRI, and neurophysiological recordings of activity from single units and local field potentials), and secondly, to facilitate re-parcellation of connectivity data according to alternative (existing or future) delineations of cortical areas and their boundaries. Thirdly, these approaches can potentially address the inner topographic heterogeneities within parcels, as suggested previously (Borra et al., 2019; Gerbella et al., 2013; ; Van Essen et al., 2018).

Fig. 5 illustrates key steps in the contours-to-atlas method. For each histological section analyzed, the location of each retrogradely labeled neuron is plotted in relation to pial, gray/white, and layer 4 contours (panel A). The surface contours of the histology section (panel B) are matched to a corresponding atlas MRI slice (panel C) based on shape similarities and regularity of spacing. The histology section contours and labeled neurons (3 separate tracers in this case) are warped to the MRI atlas sections, and the volume density of labeled neurons is calculated for each tracer injection – in this case, three separate tracers in different frontal lobe areas (panels D – F). The densities are then projected onto the atlas mid-thickness surface model and displayed as log2 (panels G – I).

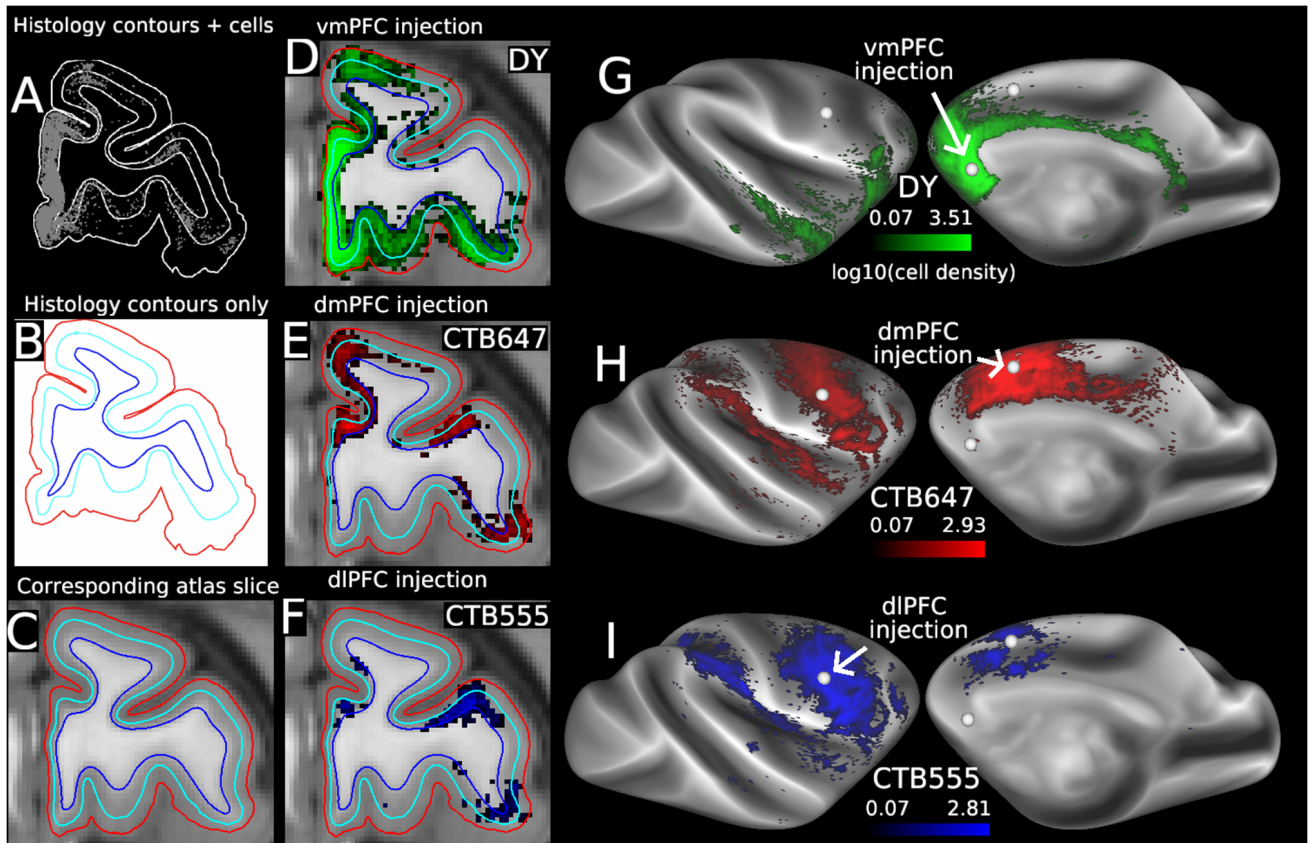

**Figure 5. Key steps in mapping retrograde tracer injection data from individual histological sections to the macaque Yerkes19 surface-based and MRI-based atlas.** A – C: matching histological section contours to corresponding atlas surface contours. In panels B-F, pial contour is red, layer 4 contour is cyan, and gray/white contour is blue. D – F. Volume density (log10 plot) of labeled neurons for each of 3 tracer injections. G – I. Dense surface maps of cell density (log10 plot) for each injection. Color bars apply to volume as well as surface maps. Cell densities thresholded at 0.1 in D-I to compensate for slight smoothing in preprocessing steps.

*Large-scale models of connectivity*. Besides the parcellated and parcellation-free analyses of connection strengths for each injected area, additional graph theoretic quantitative tracer-based analyses are currently underway. This work addresses large-scale models of the cortex and reveals such features as the core-periphery organization (Markov et al., 2013). This allows comparing the resulting large-scale models to be compared to models based on imaging approaches. In the case of dMRI tractography this approach could constitute a high-level validation procedure. The large-scale models of the cortex obtained from tract tracing are highly informative when combined with other invasive approaches. In the case of cortical electrophysiology in NHP (Bastos et al., 2015; Markov et al., 2014b) and human (Michalareas et al., 2016) , this has led to insights into the functional hierarchy of the cortex and the development of a novel model of hierarchical processing termed the dual counterstream architecture (Vezoli et al., 2020) . When these networks are considered in relation to macroscopic cortical gradients including gene expression and cell type, this approach contributes to the development of computational modeling of cortical dynamics (Wang et al., 2020).

**3.2. A macaque subcortical atlas and weighted subcortical-cortical connectivity matrix.**

Each retrograde tracer injection case contains labeled neurons in up to 30 subcortical nuclei as well as the extensive cortico-cortical connectivity patterns discussed in the preceding section. However, quantitative analyses of the subcortical connectivity patterns are hindered by the lack of an atlas that accurately and fully delineates the boundaries between all identified nuclei. We propose to generate such an atlas plus an associated connectivity database that provides counts of neurons in individual subcortical structures down to and including the brain stem. To this end, we have sectioned a macaque brain (600 coronal sections at 40 microns) and processed adjacent sections for parvalbumin, calbindin, SMI-32, NeuN and Nissl (Cresyl Violet). Using these stains in combination allows histological identification of borders between nuclei throughout. Ongoing work has established boundaries of as many as 50 subcortical nuclei that have been delineated on sections at 0.48 mm intervals. These atlas sections will be aligned to high-resolution structural MRI (spatial resolution of 0.5mm) obtained in vivo from the same animal that will in turn be aligned to the Yerkes19 atlas.

### 3.3. The claustrum as the hub of the cortex.

Of particular interest to us is the claustrum - its parcellation and cortical connectivity, and role as a possible hub in network models. Many network models of structural and functional connectivity emphasize features such as hubs and communities. The overarching expectation is that such analyses provide insight into mechanisms of integrative neuronal processes (Sporns, 2014). For example, putative hubs based on centrality measures have been reported in anterior and posterior cingulate cortex, insula, frontal, temporal and parietal cortex (Gong et al., 2009; Hagmann et al., 2008; van den Heuvel and Sporns, 2013). Hubs that are more interconnected with one another than statistically predicted constitute a subnetwork referred to as a 'rich club' (Colizza et al., 2006). Putative rich clubs in humans (van den Heuvel and Sporns, 2011) and macaque (Harriger et al., 2012) are hypothesized to play a privileged role in orchestrating interactions of the cortex.

The high density of the macaque inter-areal networks revealed by anatomical tracers calls the existence of hubs in general and rich clubs in particular into question. A formal analysis of the Markov et al. (2014a) data revealed no evidence of the existence of a cortical rich club in the macaque (Knoblauch et al., 2016). On the other hand, analyses of input weights to visual areas indicate that the claustrum has systematically high-weight connectivity exceeding that of most cortical pathways (Markov *et al.*, 2011). Our recent unpublished analyses of connectivity of 52 areas indicate that: (i) the claustrum projects to every cortical area, and (ii) claustrum inputs equal or exceed those from the thalamus to each cortical area. Developmentally, the claustrum and insular cortex originate from the same progenitor pool, and the claustrum may be considered as a persisting subplate structure related to the insula (Bruguier et al., 2020; Puelles, 2014). An intriguing hypothesis is that the claustrum constitutes a unique cortical hub that helps orchestrate large-scale neural activity patterns across the cortex. We plan to continue investigating the connectivity and functional role of the claustrum and its interactions with other structures, in particular the hippocampal formation. We will map the claustral cells from 52 injections to further investigate this hypothesis and determine FLNe values of projections of cortex to claustrum using viral tracing (Nassi et al., 2015).

### 3.4. Large-scale models of the interareal cortical network and hierarchy

The Felleman and Van Essen (1991) visual cortical hierarchy was one of the earliest connectomic large-scale models of the cortex. It built on the distinction between feedforward vs feedback connections related to the laminar origins and terminations of inter-area connections (Maunsell and van Essen, 1983; Rockland and Pandya, 1979). The Felleman and Van Essen study used pairwise comparisons of feedforward, feedback, and 'lateral' patterns to rank 32 visual areas in one of 10 levels areas in a distributed hierarchy. However, it was not a unique solution to the problem. Subsequent studies have proposed important alternative ways to analyze putative hierarchical relationships in the macaque (Barone et al., 2000; Markov et al., 2014b) and mouse (D'Souza et al., 2016; Harris et al., 2019) using laminar connectivity patterns and in humans using surrogate neuroimaging measures (Badre and D'Esposito, 2007; Demirtaş et al., 2019; Koechlin et al., 2003). We plan extensive additional analyses that are especially focused on hierarchical relationships involving prefrontal cortex.

*Laminar analyses and hierarchical organization.* An objective hierarchical distance measure reveals high structural regularity in the visual cortical hierarchy (Markov et al., 2014b). This did not generalize to prefrontal areas (Goulas et al., 2014), perhaps because the analysis was hampered by the limited number of injected areas available at that time. We will revisit this issue using data from a much larger number of prefrontal and rim area injections (see Section 3.1). We will incorporate laminar connectivity data for connections involving entorhinal cortex and other rim cortical areas that relay information between neocortex and the hippocampus, coupled with mathematical modeling of the laminar relations to the hippocampal complex. We anticipate obtaining a better understanding of where and how prefrontal areas fit into a global hierarchical network.

The structural hierarchy of visual cortex also provides a key template for analyzing inter-areal oscillatory synchrony within a functional hierarchy in macaque (Bastos et al., 2015) and human cortex (Michalareas et al., 2016). We anticipate that our analyses of structural hierarchy of the frontal cortex will provide a corresponding template underlying the functional hierarchy of the prefrontal cortex. This is highly relevant to potential clinical applications. Most psychiatric disorders implicate systems underlying cognitive functions and executive control of behavior, with the prefrontal cortex at its core. Elucidating circuit mechanisms of hierarchical processes in a Bayesian inference framework underlies the emerging field of Computational Psychiatry, which may provide a solid biological foundation for diagnosis and therapeutic treatment of mental illness (Stephan et al., 2016).

### 3.5 Imputing connectivity.

Some connectivity databases, such as an influential connectome study of the mouse (Oh et al., 2014), use computational methods to infer connectivity for cortical areas that did not receive a precisely restricted tracer injection or for other technical reasons. A robust imputation method offers the prospect of predicting the strength of missing links and better approximating a weighted structural connectome of the macaque brain. Our approach is informed by recent studies that characterize connection weight patterns in mesoscale connectomes (Horvát et al., 2016; Markov et al., 2011; Markov et al., 2014a; Theodoni et al., 2020). Thus, inter-areal networks are not random graphs, but complex networks with distinctive structural features. If weight-distance relations of areal connectivity are predictive, could injection in a subset of areas be used to impute the connectivity of the full matrix? I propose to replace this by: "Zoltan Toroczkai and his students in collaboration with Henry Kennedy lab has developed a machine learning framework that shows the high predictability of macaque cortico-cortical pathways in the macaque (Molnar et al., 2020).

### 4. Bridging neuroimaging, neuroanatomy and beyond

### 4.1. Validating neuroimaging connectivity with ground truth anatomical connectivity.

A major opportunity for NHP research is to provide a platform for using quantitative measures of 'ground truth' tracer-based connectivity to evaluate neuroimaging-based estimates of structural connectivity from tractography and diffusion MRI (dMRI) and of functional connectivity from resting-state MRI. This comparison is particularly important because these neuroimaging-based methods are very widely used in human studies, where neuroanatomical tracer-based analyses are not feasible, but they are each highly indirect methods with major biases and uncertainties. We illustrate the current state-of-the-art using examples of a parcellated analysis of tracers vs diffusion tractography and a dense analysis of tracers vs functional connectivity in the macaque.

The left panel in Fig. 6A shows the 51×51 parcellated tracer-based connectivity matrix that the Kennedy lab has been systematically collecting and analyzing in macaque since the publication of Markov et al., (2014a). Markov et al., (2014a) included 29 retrograde-tracer injections, since then the number of injections has been increased to 51 cortical seed locations, where the connectivity weights in the target parcel (91 parcels over a single hemisphere) were measured by Fraction of extrinsically Labelled Neurons (FLNe). A 51×51 matrix includes all the bidirectional connectivity weights, scaled by the total weights of the connectivity and symmetrized by averaging weights of bidirectional connectivity. The middle and right panels of Fig. 6A show the diffusion tractography matrices in *ex vivo* (DTe) high-spatial resolution dMRI (Donahue et al 2016) (N=1) and *in vivo* (DTi) HCP-style high-angular resolution dMRI (Autio et al., 2020b) (N=15, average), respectively, both represented by Fractional Scaling of streamlines (FSe) considered to correspond to FLNe (Donahue et al., 2016). Fig. 6B shows exemplar F5-seed parcellated connectivity maps for tracer (left), *ex vivo* (middle) and *in vivo* diffusion tractography (right), showing similar but far from identical connection weights. A systematic scatterplot comparison of 51×51 matrices between tractography FSe vs tracer FLNe values using Spearman's correlation coefficient revealed a moderate correlation (Spearman's rho = 0.56 and 0.60 for true-positive DTe and DTi respectively) (Fig. 6C, D) similar to values observed with the 29×29 matrix (Donahue et al 2016; Autio et al., 2020b). The color scheme in Fig 6C and D indicates the connection distance, showing that the correlation between tractography and tract tracing decreases with the increasing connection distance. However, the distance of connections is estimated by group-averaged streamline length of tractography whose reliability still needs further study. Both the DTe and DTi showed high sensitivity (99% and 100%, respectively, at a threshold of $10^{-6}$), which suggests that HCP-style in vivo NHP data acquisition can provide connectivity estimates comparable in fidelity to the higher resolution attainable with postmortem scans. But both approaches need significant improvement particularly in specificity (0.8% and 0.4%, respectively) to better estimate ground-truth anatomical connectivity. Correlation based on true positives reveals the agreement of the connection weight between tractography and tract tracing for the connections detected by tract tracing, but it is important to also consider false negatives and false positives in a receiver operator curve type analysis (Donahue et al 2016). Also note that Fig 6C and D show an orthogonal regression that accounts for the variance in empirical values of TT and DTe or DTi, thus, we may need to evaluate whether the refinement of tracer mapping ('dense mapping' and 're-parcellating') described in section 3.1 can further improve linear fitting. We are mapping tracer and neuroimaging-based connectivity in the same animal in order to resolve cross-subject variability in the cortical connectivity (Markov et al., 2011). We also hope to facilitate algorithm development and evaluation of *in vivo* tractography by making the datasets publicly available. For example, one potential avenue is to invoke geometric constraints on the trajectories of neighboring fiber bundles in gyral blades allowing unbiased sensitivity over the surface and reliable dense diffusion connectivity.

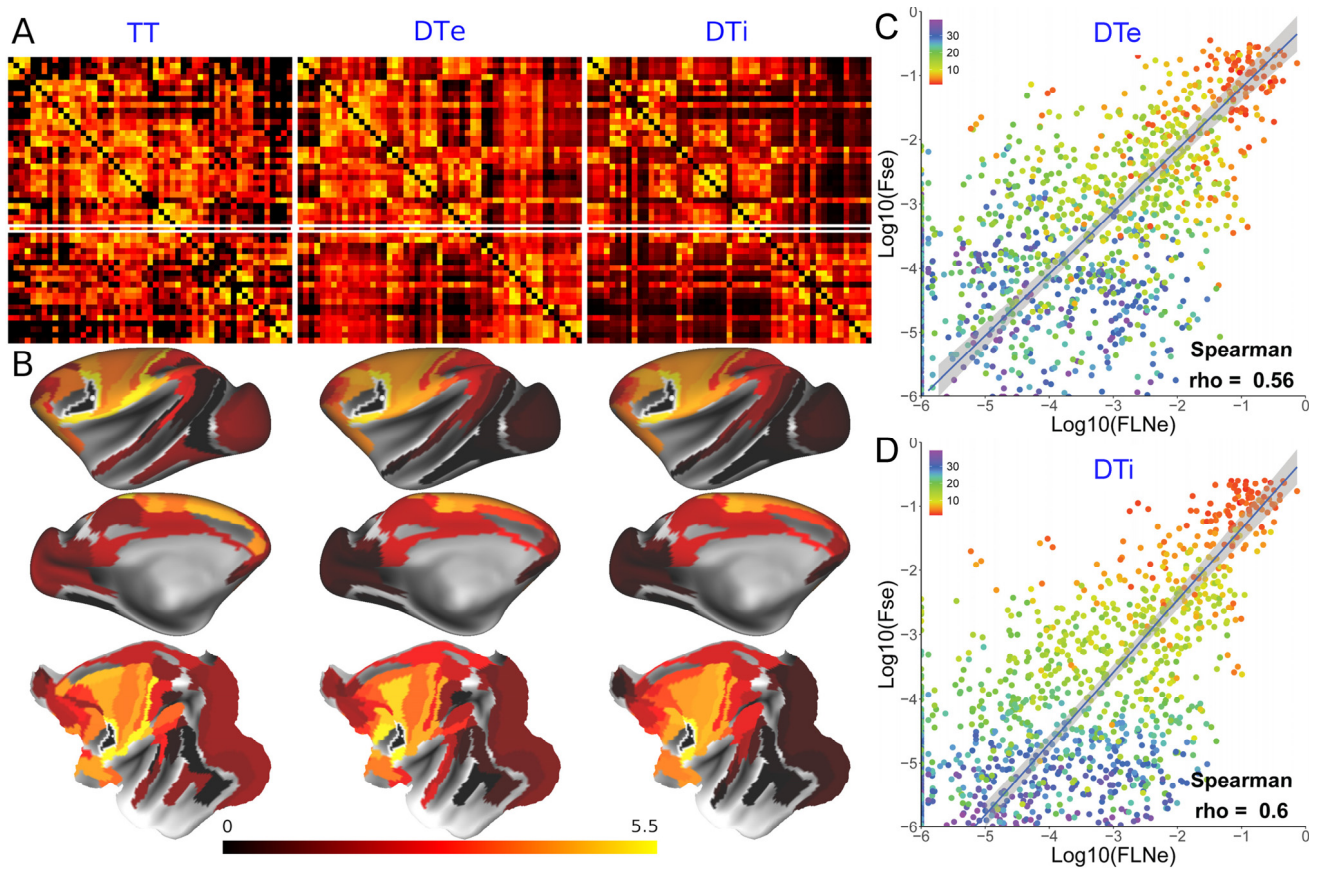

**Figure 6. Cortico-cortical connectivity based on tract tracing and *ex vivo* and *in vivo* diffusion tractography**
**A**). Connectivity matrices of 51 areas revealed by tract tracing and tractography. Left: 51x 51 matrix of bidirectional tract tracing (TT) shown retrograde neuronal tracer connectivity injected in 51 cortical areas, and analyzed over 91 cortical areas, shown as FLNe by log10 scale (Markov et al 2014a), middle: *ex vivo* diffusion connectivity weights (DTe) presented as log 10 scale of FSe (N=1) (Donahue et al., 2016), right: *In vivo* diffusion connectivity matrix weights (DTi) shown in FSe (N=15, averaged) (Autio et al., 2020b). The FLNe and FSe of F5 seed-connectivity are highlighted by white line. **B**) An exemplar surface maps of tracer connectivity weights for an injection in area F5 (white dot ), *ex vivo* diffusion tractography of F5 seed, and *in vivo* diffusion tractography of F5 seed. Upper: lateral surface of hemisphere, middle:

medial surface of hemisphere, bottom: flatmap. **C**) Scatter plot of TT FLNe vs DTe FSe,and **D**) TT FLNe vs DTi FSe. The points are color coded by the weighted average connectivity distance estimated from diffusion tractography (N=15) (see colormap at the upper left corner of each graph). At a threshold of $10^{-6}$, the numbers of true-positive/true-negative/false-positive/false-negative were 1030/2/235/8 and 1038/1/236/0 in DTe and DTi respectively, resulting in sensitivity of 99% and 100% and specificity of 0.8% and 0.4%, respectively. The blue line (and gray bands around the line) indicates an orthogonal line regression that accounts for the variance in empirical values of both the x- and y-axis. Note that in A and B, the FLNe and FSe are shown as actual log values plus 6 for visualization purposes using Connectome Workbench (wb_view).

Another important unresolved question is how closely functional connectivity (FC) matches tracer data. Comparison of FC with whole brain quantitative monosynaptic neural tracer data merits further study to determine to what extent FC is dependent on physically wired connections, as well as to improve the models and techniques used to calculate FC. Fig. 7 shows exemplar results from a comparison between dense connectivity from 31 tracer injections and full-correlation (normal pairwise correlation) FC from 30 macaques. Fig. 7A the tracer-vs-FC correlation is relatively high (r = 0.56) for an area 7b injection/seed, but in Fig. 7B the correlation is much lower (r = 0.06) for an area 9/46v injection. Simple FC reflects not just direct, monosynaptic connections but also indirect (polysynaptic) connections, which are important for plasticity and network reorganization (Kudoh and Shibuki, 1997; Miroschnikow et al., 2018; Sakurai, 2016). Partial correlation has been hypothesized to reduce sensitivity to indirect connectivity, and therefore in theory should agree with tracer data better than full correlation. However, in our preliminary analysis, partial correlation performed slightly worse than full correlation on average (r = 0.39 vs 0.42), though optimal regularization has yet to be determined. Parcellated analysis can also provide higher fidelity by reducing the effect of noise on fMRI signals. Other models taking into account nonlinear and dynamic network interactions also merit testing (Vidaurre et al., 2017). Correspondence of contralateral hemispheric connections is also worth further investigation.

Increasing and evaluating sensitivity/specificity/reproducibility of the measurement systems (including tracer and MRI) are important for understanding not only the 'ground truth' of NHP species but also variability between subjects (see also Autio et al., 2020a). Cross subject variability may be validated by an approach for multi-modal evaluations (e.g. tracer and dMRI both or tracer and fMRI both) in the same subject. There is an individual fingerprint inter-area connectivity, so that repeat injections in the same individual lead to near identical connectivity profiles, whereas across individual repeats show variability. Hence, if tractography and tract tracing are reflected connectivity patterns, then one would predict they would show an improved correlation when they are carried out in the same brain.

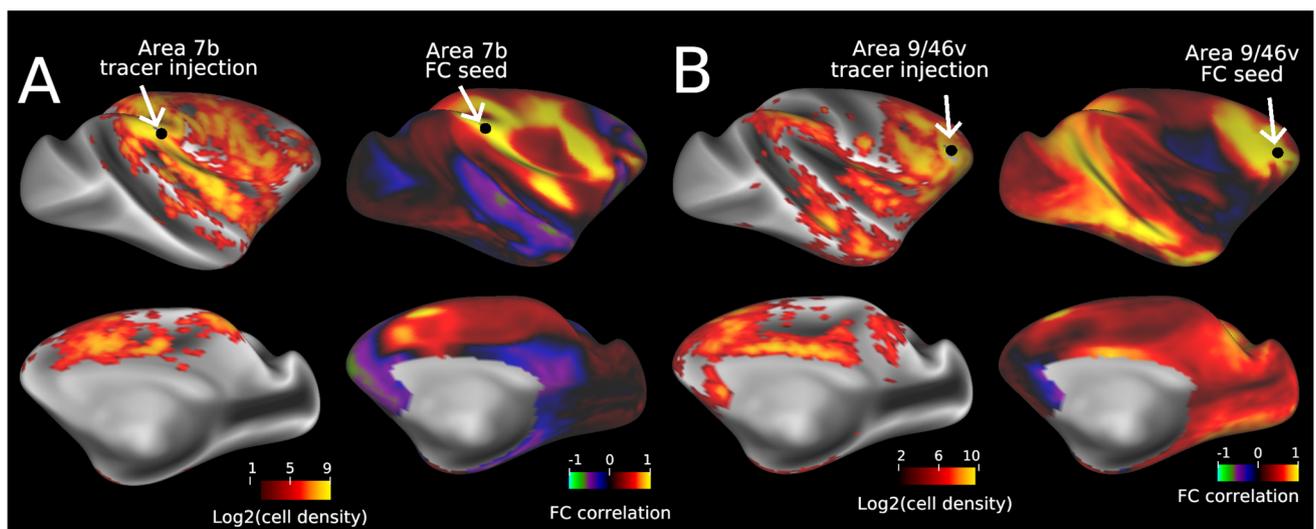

**Figure. 7.** Comparing dense tracer connectivity with functional connectivity
Two exemplar injection sites (**A**, area 7b and **B**, area 9/46v) and corresponding tracer connectivity and seed location from a dataset involving 31 tracer injections (left in each panel) and functional connectivity and the seed from the initial NHP_NNP fMRI dataset, Mac30BS (right in each panel). The Mac30BS dense connectivity was generated using thirty sedated macaques with 102-min fMRI BOLD scan duration each.

## 4.2. Towards a 21st century macaque brain atlas - multimodal parcellation including histology and neuroimaging

*A perspective on cortical parcellation.* An important goal of NHP studies is accurate parcellation of the cerebral cortex. In the late 20th century, the macaque was one of the most intensively studied laboratory animals and was arguably among the best understood, by virtue of analyses that made use of the 'FACT' strategy of using information about Function, Architecture, Connectivity, and Topography (FVE91). In recent decades, accurate parcellation of mouse and human cortex has been impressive and leapfrogged the macaque and marmosets in terms of how closely it likely approaches a ground-truth parcellation (Van Essen and Glasser, 2018). For human cortex, key advances enabling identification of the 180 areas in the HCP_MMP1.0 parcellation (Glasser et al., 2016a) included (i) the acquisition of vast amounts of exceptionally high quality multimodal neuroimaging data from healthy young adults as part of the HCP (Van Essen et al., 2012c); (ii) systematic processing of each modality via HCP-style preprocessing that maximizes signal, corrects artifacts and minimizes spatial blurring, and maximizes accurate alignment of functionally corresponding regions across subjects (Glasser et al., 2016b); and (iii) a parcellation strategy that identified candidate areal boundaries using a 'gradient-ridge' approach that demanded consistency across multiple modalities spanning the FACT domains together with an areal classifier that takes advantage of all the information in a high-dimensional feature space to identify each area in each individual. Ideally, this type of neuroimaging-based parcellation strategy would be combined with one based on postmortem architectonics, including cytoarchitecture, chemoarchitecture (Zilles and Amunts, 2009), cells (Murakami et al., 2018; Ren et al., 2019), gene expression (Shimogori et al., 2018; Yamamori and Rockland, 2006) and other modalities, but such a fusion across in vivo and ex vivo datasets has yet to occur. For the mouse, the success of two separate teams in identifying very similar 41-area parcellations (Gămănuţ et al., 2018; Harris et al., 2019) was based mainly on postmortem architectonics using multiple markers (and tangentially-sliced cortical flat mounts) plus additional information about connectivity and topographic organization.

There are daunting challenges in applying these lessons to the macaque and generating a cortical parcellation that approaches ground-truth and exploits use of both invasive and non-invasive data. Here, we outline our general thoughts and observations on what it will likely take to succeed. In some respects, the macaque and marmoset are 'caught in the middle'. Their brains are not large enough to yield neuroimaging data with sufficient spatial resolution and the necessary signal-to-noise values. It is too hard to collect big NHP data in awake conditions. This means that given current 3T scanner technology in NHP, it is not possible to simply emulate the success of the HCP's 'neuroimaging-only' approach. On the other hand, NHP brains are too large, the individual variability is too great, and the functional organization is too complicated to easily emulate the microscopic architectonic patterns that are evident in tangentially sliced mouse flatmaps.

We think the optimal strategy for an accurate macaque cortical parcellation will entail a hybrid approach that combines (i) ultra-high resolution in vivo and postmortem neuroimaging, (ii) postmortem multimodal histological atlases of unprecedented spatial resolution and alignment, and (iii) a multimodal parcellation strategy that focuses on consistent spatial gradients and high-dimensional feature maps for areal classification. For postmortem architectonics, it may be critical to carry out layer-specific analyses of spatial gradients and feature maps, given the major differences in architecture across layers. This will demand alignment fidelity of ~50 um or better throughout the brain. This may be attainable using promising approaches such as the VISoR method (Wang et al., 2019) that may facilitate high-throughput, high-resolution optical imaging of thick tissue slabs that can be accurately aligned across slabs and repeatedly stained using multiple immunohistochemical markers (e.g., parvalbumin, somatostatin and neurofilament protein SMI-32).

For the ultra-high resolution in-vivo neuroimaging, ultra-high field MRI scanners have a great potential by combining parallel imaging hardwares and sequences for high-contrast & high-speed fMRI in NHP. Ultra-high resolution fMRI is expected to resolve cortical layer-level functional organization (Huber et al., 2020). Layer-dependent whole brain connectomes along with tract tracing in NHP would deepen our understanding of the functional and anatomical correlates of feedforward/feedback directionality and hierarchy in this species (Felleman and Van Essen, 1991; Rockland and Pandya, 1979). Beginning in 2020, Essa Yacoub's lab in UMinn has joined this project, as his lab has recently applied a ultra-high field (10.5T) MRI scanner for NHP study, and established an ultra-high resolution fMRI in macaque monkeys (Yacoub et al. this issue). The resolution of the fMRI data sampling (isotropic voxel of 0.75mm) is close to one third of the median cortical thickness and the field of view (410 mL) fully covers the whole brain of this species (see Table 1) with high signal-to-noise ratio (Yacoub et al., this issue). The surface-based layer analysis is now available for human ultra-high field studies (Polimeri et al., 2010), where whole brain high-resolution fMRI at 2 mm (74% of median cortical thickness, mCT) partly separated signals between two layers, and fMRI with a higher resolution > 1.6mm (59% mCT) more reasonably separated signals by simulation (Coalson et al., 2018). By applying the same simulations, we explored how our approach for spatial resolution distinguishes the signals from different layers of superficial and deep. Fig. 8 shows the results of simulations: while conventional resolution in macaque fMRI (2 mm, 95% of mCT) cannot separate the signals from two layers, fMRI protocols in NHP_NNP partially separated at 3T (1.25 mm, 59% mCT ) and largely separated signals from two layers at ultra-high field MRI (0.75 mm, 33% mCT).These preliminary data suggest the ultra-high resolution fMRI at ultra-high field MRI is potentially useful to discriminate signals from dichotomized layers, so optimization of scanning protocols and analysis, collection of multiple subjects data will be done to investigate the layer-dependent functional connectome.

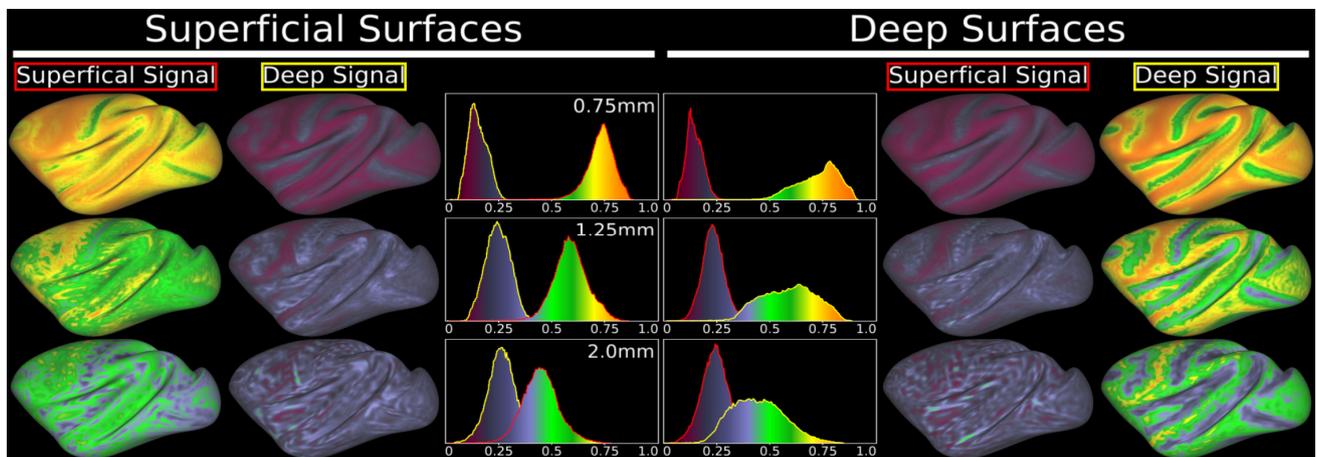

**Figure 8.** Results of simulation for partial volume/geometric effects in 'layer' fMRI analysis at various resolutions in macaque cerebral cortex.

Simulation performed estimation of 'layer 4' surface for each subject of macaques (N=30) from Mac30BS, created fMRI-resolution volumes of simulated superficial and deep cortical signal (above and below layer 4, respectively), and re-mapped each layer's signal onto the subject's surfaces with ribbon mapping (pial to layer4, and layer4 to white), and these values were then averaged across subjects. In the corresponding layer/surface, a value closer to one indicates better response to layer-specific signal, where the values for the other layer indicate 'spill-in' of signals to the undesired layer. The layer 4 surface is approximated by the 'equal volume method' (locally, the same amount of cortical volume is superficial to the new surface as is deeper than it) (Van Essen and Maunsell, 1980). The use of conventional resolution in NHP (=2.0mm) results in significant loss or spill-out of signals in the corresponding layer, and spill-in into the other layer resulting in the overlap of histograms. NHP_NNP 3T protocol (1.25mm) was partly overlapped in histogram, and ultra-high field MRI protocol (0.75mm) showed distinct separation between peaks in the histograms, suggesting differentiation of signal between laminae in surface-based group analysis if one only considers geometric constraints (note that the separability of lamina also depends on the point-spread-function of BOLD fMRI or other fMRI approaches). See also Fig. S8 in (Coalson et al., 2018). Data at https://balsa.wustl.edu/0LMl2

### 4.3. Marmosets as an NHP model system.

There is increasing interest in using marmosets in the field of neuroscience, as well as connecomics by collecting and sharing large amounts of data on cellular connectivity (Majka et al., 2020) and high-quality diffusion tractography (Liu et al., 2020). Marmosets provide an important bridge across the gap in scale and technology (genetics, histology vs MRI) and evolutionary neurobiology (rodents vs monkeys vs human). Many cortical areas involved in higher cognitive function in humans may lack evolutionary homologues in rodents (Preuss, 2000; Van Essen and Glasser, 2018). The marmoset offers several advantages as a model system; (i) the marmoset brain shares some aspects of the developmental process and anatomical structure of the human brain (Homman-Ludiye and Bourne, 2017); (ii) the marmoset has similar social behaviours to humans, including particularly a strong relationship between parents and offspring (Saito, 2015); (iii) the marmoset displays unique social vocal communication and there is a likely convergent evolution in this characteristic (Eliades and Wang, 2008); (iv) several neurological disease models of marmoset are analogous to human disease; (v) some higher cognitive tasks in marmosets are equivalent to those found in macaques; (vi) the marmoset can be handled with comparative ease owing to its small body size; and (vii) the marmoset has a strong reproductive efficiency, is suitable for understanding genetic effect of brain evolution, behavior and diseases (Izpisua Belmonte et al., 2015; Sasaki et al., 2009). Recently, Majka et al. (Majka et al., 2020) revealed 143 injections of retrograde tracers in 59 marmosets and the distribution of the tracers were demonstrated in a 116-area parcellation in volume space, which may be useful to compare with and/or validate non-invasive connectivity data. Gene expression patterns in the marmoset brain have also been reported recently (Shimogori et al., 2018) and may merit multi-modal atlasing. Therefore, connectome studies in marmosets and comparisons with other NHP should also benefit from future refinements in standardized mapping of multimodal data.

The downside of our 'all species on the same scanner' approach may be limited spatial resolution for the small NHP like marmoset. We determined it based on cortical thickness, thus particularly in this species may not achieve the comparable number of sampling over the cortical surface or brain volume as other species, since the isometric difference between human and marmoset is 1.7-fold in cortical thickness whereas it is 10-fold and 5-fold in surface and brain volume, respectively (Table 1). The spatial resolution in our approach also depends on the limits allowed by the sequence, at which we also configured comparable temporal resolution of fMRI across species (~0.7 sec). Smaller sampling over the surface may be mitigated by an expected smaller number of cortical areas in the marmoset than higher primates (see Table 1). That said, an alternative approach may be 'marmoset on the ultra-high field scanner' to take advantage of spatial resolution (Hori et al., 2020; Liu et al., 2019), but further refinements may be needed to overcome impediments in parallel imaging hardwares and softwares for achieving comparable temporal resolution, as well as in high-quality 3D structural imaging for successful cortical surface-based analysis and comparable myelin mapping across NHPs (see section 2.2).

### 4.4. Translating neuroimaging to social behaviors and genes in NHP

One unsolved issue is the extent to which NHP models will allow us to study brain diseases such as psychiatric diseases. In particular, an outstanding question is the individual variability and associations of brain organization, social behaviors and genetics in NHPs (Izpisua Belmonte et al., 2015), as has been explored for human developmental and psychiatric diseases (Van Essen et al., 2006; Wei et al., 2019) and genome-connectome studies (Elliott et al., 2018; Miranda-Dominguez et al., 2018). Vasopressin and oxytocin are neuropeptides implicated in the development and maintenance of social behaviors in mammals (Donaldson and Young, 2008; Meyer-Lindenberg et al., 2011). Recently, we showed that marmosets' prosocial behaviors and individual variability are associated with the vasopressin V1a receptor gene (*AVPR1A*) and other genetic polymorphisms related to oxytocin and dopamine transmission (Inoue-Murayama et al., 2018; Weiss et al., 2020). Based on these preliminary data, the NHP_NNP project plans to investigate the association of the neuroimaging-based brain connectome with genetic polymorphism and social behaviors.

For the social behaviors of marmosets, we plan to perform reliable assessments by a well-established personality rating scale, Hominoid Personality Questionnaire (Weiss, 2017; Weiss et al., 2009), which proved to be reproducibly correlated with other biological markers and genotyping in marmosets (Inoue-Murayama et al., 2018; Weiss et al., 2020). The reliability of this behavioral analysis is based on long-term housing and observations (>one year) by the animal keepers. For macaque in which most of the data is collected in a short stay (less than 3 weeks), we plan to analyze 'gaze-sensitivity', which is known to be a hallmark of adaptive, social behaviors interacting with others (Davidson et al., 2014). For genotyping in macaques and marmosets, buccal swabs are obtained from each animal and analyzed as described previously (Inoue-Murayama et al., 2018), in which the target DNA fragments were chosen from those that are associated with neurotransmission of dopamine, serotonin, oxytocin and vasopressin as previously (Inoue-Murayama et al., 2010, p.; Staes et al., 2016). Hair samples are also obtained for analyzing the cortisol level as a marker of chronic stress (Inoue-Murayama et al., 2018; Yamanashi et al., 2016).

Correlating behavior and genetics with brain connectomics is challenging due to small sample size in NHP studies as has been demonstrated in human neuroimaging-genetic studies (de Zubicaray et al., 2008). Studies in twins may be a powerful approach to investigate contribution of genetics to brain connectome and/or behaviours, as done in humans (Peper et al., 2007), but of course that only allows for the investigation of heritability, and not specific gene associations. That said, a recent study analyzing a total of 142 chimpanzees also encouragingly revealed a relationship of polymorphism of *AVPR1A* to the size of the association cortices (Mulholland et al., 2020).

### 4.5. Sharing neuroimaging data, neuroanatomical atlases and connectivity data.

Given the scope of these challenges combined with the large potential payoff to the field of neuroscience, the prospects for success may be highest with a large-scale international effort that would engage the expertise of many investigators and multiple institutions. The standardization and harmonization of the data acquisition, analysis, and sharing are important features of the approach. An international alliance to tackle major outstanding questions, like PRIME-DE, may be one promising solution for future brain science using NHPs (Milham et al., 2018).

The neuroimaging data, neuroanatomical atlases and connectivity data generated by the current project described above will be valuable resources for the NHP (and broader neuroscience) research community. The results of the neuroimaging studies generated by this collaboration will be made publically available from BALSA (https://balsa.wustl.edu/). Unprocessed data will be made publicly available through PRIME-DE (http://fcon_1000.projects.nitrc.org/indi/indiPRIME.html) after publication. The preprocessing pipeline, HCP-NHP pipeline, is currently maintained at an independent github repository (https://github.com/Washington-University/NHPPipelines), but will be merged into the main HCP Pipelines repository for ease of maintenance in the future. The imaging and anesthesia protocols are available at RIKEN Brain/MINDS-beyond site (https://brainminds-beyond.riken.jp/hcp-nhp-protocol).

The neuroanatomical atlases and connectivity data from individual cases in their native configuration will also be shared as much as is feasible (i.e., relative to histological section contours) and after registration to the Yerkes19 atlas. Data sharing will be via Core-Nets (https://core-nets.org/) and the BALSA database (https://balsa.wustl.edu), and will use 'scene files' in Connectome Workbench format as the primary way of organizing each dataset (Van Essen et al., 2017). Indeed, some of the datasets presented here are accessible via https://balsa.wustl.edu/study/show/Klr0B Online access to the unparcellated datasets will allow others to re-parcellate the data according to their preferred criteria.

We will also share the genetics and behavioral data at the RIKEN Brain/MINDS-beyond site (https://brainminds-beyond.riken.jp) along with neuroimaging data to accelerate multi-site collaboration for larger samples of NHPs, as is being performed in ongoing human neuroimaging projects such as the ABCD (Casey et al., 2018) and Brain/MINDS-beyond (Koike et al., 2020).

### 5. Conclusion

In addition to providing an objective validation of imaging techniques and a deeper understanding of their significance, the findings of the NHP_NNP will be relevant for a broad range of issues. How unique is the human brain, and how does it differ from other primates? While the mouse has become the most widely used model organism in neuroscience, there is increasing appreciation of the critical importance of NHP studies owing to their closer evolutionary proximity to humans. Comparison of the organizational principles in mouse and NHP will address the complementary ways in which the human brain can be modeled by work in rodents vs NHPs in investigation of neural systems and their disorders. Our freely shared brain-wide connectivity maps and their relationship to surface registered imaging will be of great utility for researchers, for example, looking at large-scale electrophysiology and gene expression networks, as well as much needed dynamic models of cortical structure and function.

A major incentive to investigate nonhuman primates is that many human neurological and neuropsychiatric diseases are inadequately modeled in rodents due to numerous primate-specific features primarily stemming from divergent evolution (Bakken et al., 2016; Smart et al., 2002). In this respect there is increasing evidence of the importance of human specific genes and gene regulation, which recently have been shown to target the supragranular layers (Heide et al., 2020; Won et al., 2019). These findings open up the possibility that interareal communication is a target for primate evolutionary adaptation, which is further supported by the increased heterogeneity of glutamatergic cell types in terms of morphology, electrophysiology and gene expression observed in the supragranular layers of the human cortex(Berg et al., 2020; Dehay and Kennedy, 2020). Because the network features of the cortex using both imaging and tract tracing have shown marked correlations with transcriptomic specialization (Burt et al., 2018), we expect that comparative investigations of these features across primate species, but also in different pathologies and their models, will become an increasingly important area of research, calling for the levels of standardization that we advocate here.

**Acknowledgements**
This research is partially supported by the program for Brain/MINDS-beyond from Japan Agency for Medical Research and development, AMED (JP20dm0307006) (T.H.), NIH F30 MH097312 (M.F.G.), RO1 MH-060974 (D.C.V.E.), RF1 MH116978 (E.Y.), LABEX CORTEX (ANR-11-LABX-0042) of Université de Lyon (ANR-11-IDEX-0007) operated by the French National Research Agency (ANR) (H. K.), ANR-11-BSV4-501, CORE-NETS (H.K.), ANR-14-CE13-0033, ARCHI-CORE (H.K.), ANR-15-CE32-0016, CORNET (H.K.), Chinese Academy of Sciences President's International Fellowship Initiative. Grant No. 2018VBA0011 (H.K.), ANR-19-CE37-025, DUAL_STREAMS (K.K.) and KAKENHI 19H04904 (M.M.). All the authors have no conflicts of interest to declare.